\documentclass[12pt]{article}

\usepackage{amsmath,amssymb,graphicx} 
\usepackage{epic}
\usepackage{epsfig}

\newcommand{\del}{\partial}
\newcommand{\ov}{\overline}
\newcommand{\drawsquare}[2]{\hbox{%
\rule{#2pt}{#1pt}\hskip-#2pt
\rule{#1pt}{#2pt}\hskip-#1pt
\rule[#1pt]{#1pt}{#2pt}}\rule[#1pt]{#2pt}{#2pt}\hskip-#2pt
\rule{#2pt}{#1pt}}
\newcommand{\Yfund}{\raisebox{-.5pt}{\drawsquare{6.5}{0.4}}}

\newcommand{\beq}{\begin{eqnarray}}
\newcommand{\eeq}{\end{eqnarray}}
\newcommand{\centeron}[2]{{\setbox0=\hbox{#1}\setbox1=\hbox{#2}\ifdim
                           \wd1>\wd0\kern.5\wd1\kern-.5\wd0\fi \copy0
                           \kern-.5\wd0\kern-.5\wd1\copy1\ifdim\wd0>\wd1
                           \kern.5\wd0\kern-.5\wd1\fi}}
\newcommand{\ltap}{\>\centeron{\raise.35ex\hbox{$<$}}
                   {\lower.65ex\hbox{$\sim$}}\>}
\newcommand{\gtap}{\>\centeron{\raise.35ex\hbox{$>$}}
                   {\lower.65ex\hbox{$\sim$}}\>}

\newcommand\ZZ{\hbox{\zfont Z\kern-.4emZ}}
\font\zfont = cmss10 

\newcommand{\refcite}{\cite}

\begin{document}

\title{\bf 2009 TASI Lecture -- Introduction to Extra Dimensions
\vspace{0.3in}}

\author{\textbf{Hsin-Chia Cheng}\vspace{0.2in}
\\
\normalsize{\it Department of Physics, University of California}\\
\normalsize{\it Davis, California 95616, USA}}

\date{}

\maketitle
\vspace{0.5in}
\begin{abstract}
We give a brief introduction to theories with extra dimensions. We first introduce the basic formalism for studying extra-dimensional theories, including the Kaluza-Klein decomposition and the effective theory for 3-branes. We then focus on two types of scenarios: large extra dimensions as a solution to the hierarchy problem and TeV$^{-1}$-size extra dimensions with Standard Model fields propagating in them. We discuss the experimental tests and constraints on these scenarios, and also the questions in particle physics which may be addressed with the help of extra dimensions. This is the write-up of the lectures given at the 2009 TASI summer school. Other interesting topics such as warped extra dimensions are covered by other lecturers. 
\end{abstract}


\newpage

\section{Introduction}
\label{sec:intro}

The idea of extra dimensions started from attempts to unify different forces in nature. In 1914 Nordstr\"om~\cite{Nordstrom} proposed a 5-dimensional (5D) vector theory to simultaneously describe electromagnetism and a scalar version of gravity. After the discovery of General Relativity, Kaluza~\cite{Kaluza} (1919) and Klein~\cite{Klein} (1926) realized that the 5D Einstein's theory with one spatial dimension compactified on a circle can describe both the 4-dimensional (4D) gravity and electromagnetism. However, the Kaluza-Klein (KK) theory has many problems and is not a viable model to describe nature.  Higher dimensional theories received renewed interests in the late 1970's and 1980's because of the developments in supergravity and superstring theories. The consistency of superstring theory requires extra dimensions. However, the extra dimensions considered then are extremely small, of the order Planck length $M_{pl}^{-1}$, which is beyond any possible experimental reach.

In the 1990's people began to consider the possibility that some extra dimensions are much larger than the Planck length:
\begin{itemize}
\item Antoniadis~\cite{Antoniadis:1990ew} (1990) proposed TeV$^{-1}$-size extra dimensions related to supersymmetry (SUSY) breaking.

\item Ho\v{r}ava and Witten~\cite{Horava:1995qa,Horava:1996ma} (1996) noticed that an extra dimension $\sim (10^{12}\; {\rm GeV})^{-1}$ in M-theory can lower the string scale to the grand unification scale $M_{\rm GUT} \sim 10^{16}$ GeV and hence can unify gravity together with other forces at the same scale.

\item The discovery of D-branes in string theory by Polchinski~\cite{Polchinski:1995mt} (1995) provides a natural setting for different fields living in different number of extra dimensions, {\it e.g.,} Standard Model (SM) fields can be described by open strings which are localized on lower-dimensional D-branes, while gravitons are described by closed strings which propagate in all dimensions.

\item The idea of extra dimensions became popular in phenomenology after Arkani-Hamed, Dimopoulos and Dvali~\cite{ArkaniHamed:1998rs} (1998) considered large extra dimensions as a solution to the hierarchy problem. The way to address the hierarchy problem with large extra dimensions is described below.

\item Warped extra dimensions (Randall and Sundrum~\cite{Randall:1999ee,Randall:1999vf}, 1999) and AdS/CFT correspondence (Maldacena~\cite{Maldacena:1997re}, 1998) provide new exciting possibilities to understand and construct models related to the weak scale. They are covered in Tony Gherghetta's lectures~\cite{Gherghetta}, so we will focus on flat extra dimensions here.
\end{itemize}

\noindent {\bf Large extra dimensions as a solution to the hierarchy problem:} In Standard Model the electroweak symmetry is broken by the vacuum expectation value (VEV) of a scalar Higgs field. However, the electroweak scale is unstable under radiative corrections as the mass-squared of a scalar field receives quadratic contributions from its interactions. The natural scale to cut off the quadratic contributions is the Planck scale when the quantum gravity effects become important. Then, the question is why electroweak scale ($\sim 100-1000$ GeV) is so much smaller than the 4D Planck scale ($\sim 10^{19}$ GeV). One can turn this question around  and ask: why is gravity so weak compared with other interactions in the Standard Model. One possibility is that the 4D Planck scale may not be a fundamental scale and the scale of quantum gravity is actually much lower if there exist large extra dimensions~\cite{ArkaniHamed:1998rs,Antoniadis:1998ig,ArkaniHamed:1998nn}.

\begin{figure}
\begin{center}
\epsfig{file=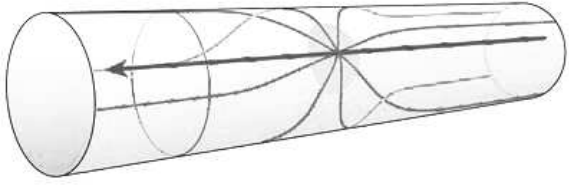,width=3.0in}
\end{center}
\caption{Force lines from a point mass in compact extra dimensions}
\label{fig:flux}
\end{figure}

Let us consider Newton's law in $4+n$ dimensions:
\begin{equation}
F(r) \sim \frac{G_N^{(4+n)} m_1 m_2}{r^{n+2}} = \frac{1}{M_{pl (4+n)}^{n+2}} \, \frac{m_1 m_2}{r^{n+2}}.
\end{equation}
If $n$ extra dimensions are compact with size $L= 2\pi R$, then the force lines from a source mass have to go parallel in extra dimensions when the distance in the usual 3 spatial dimensions is larger than $L$ (Fig.~\ref{fig:flux}),
\begin{eqnarray}
F(r) &\sim& \frac{1}{M_{pl (4+n)}^{n+2}} \, \frac{m_1 m_2}{r^{n+2}}, \quad \mbox{ for } r \ll L, \\
F(r) &\sim& \frac{1}{M_{pl (4+n)}^{n+2}} \, \frac{m_1 m_2}{L^n r^{2}}, \quad \mbox{ for } r \gg L.
\end{eqnarray}
Comparing the last expression with the 4D Newton's law,
\begin{equation}
F(r) \sim \frac{1}{M_{pl (4)}^{n+2}} \, \frac{m_1 m_2}{r^{2}},
\end{equation}
we have
\begin{equation}
M_{pl(4)}^2 \sim M_{pl(4+n)}^{n+2} L^n = M_{pl(4+n)}^{n+2} V_n,
\end{equation}
where $V_n$ is the volume of the compact extra dimensions.
If we take the fundamental scale $M_{pl(4+n)} \sim 1$ TeV and the 4D Planck scale $M_{pl(4)} \sim 10^{19}$ GeV, then we have (assuming extra dimensions have the same size)
\begin{equation}
L \sim \left( \frac{M_{pl(4)}^2}{M_{pl(4+n)}^{n+2}}\right)^{1/n} \sim 10^{32/n} \mbox{ TeV}^{-1} \sim 10^{32/n} 10^{-17} \mbox{ cm},
\end{equation}
\begin{eqnarray}
n&=&1  \Rightarrow  L \sim 10^{15} \mbox{ cm } (> 1 \mbox{ AU}), \mbox{ obviously ruled out,} \nonumber \\
n&=&2  \Rightarrow  L \sim 1 \mbox{ mm }, \mbox{ allowed in 1998, but current bound $L < 200 \, \mu$m} \nonumber \\
n&=&3  \Rightarrow  L \sim 10^{-6} \mbox{ cm }. \nonumber
\end{eqnarray}

\begin{figure}
\begin{center}
\epsfig{file=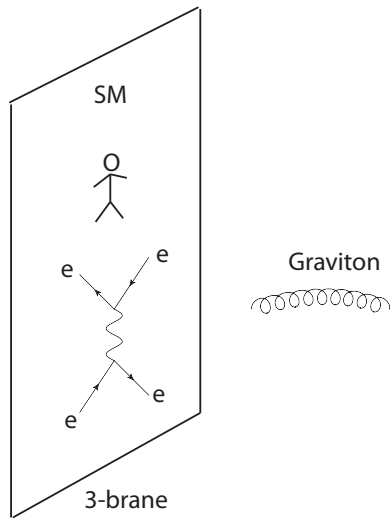,width=2.0in}
\end{center}
\caption{The brane world}
\label{fig:braneworld}
\end{figure}

On the other hand, SM has been well-tested up to a few hundred GeV to TeV, so SM field cannot propagate in extra dimensions with size $R \gtrsim 1 \mbox{ TeV}^{-1}$.  If there are large extra dimensions, SM fields have to be localized on a 3-brane as shown in Fig.~\ref{fig:braneworld} (with thickness 
$\lesssim 1 \mbox{ TeV}^{-1}$). It was surprising that such a scenario is alive and not ruled out experimentally or observationally. In Sec.~\ref{sec:exp} we discuss various constraints from high-energy and low-energy experiments as well as from astrophysics and cosmology. In the next 2 sections we will first develop the formalism for studying theories with extra dimensions.

\section{Kaluza-Klein Theory}
\label{sec:kk}

\begin{figure}
\begin{center}
\epsfig{file=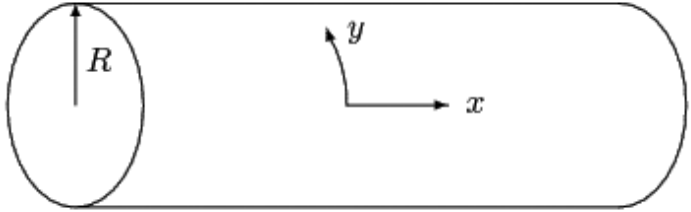,width=4.0in}
\end{center}
\caption{One extra dimension compactified on a circle}
\label{fig:tube}
\end{figure}
We first discuss how a higher-dimensional field theory reduces to 4-dimensional after the extra dimensions are compactified. As an illustration let us consider one extra dimension compactified on a circle with radius $R$ (Fig.~\ref{fig:tube}). The coordinates are denoted as $x^M = (x^\mu, y)$, where $M=0, 1, 2, 3, 5$, $\mu= 0, 1, 2, 3$ and $y=x^5$ is the coordinate in the direction of the extra dimension. The compactification means that the points $y$ and $y+2\pi R$ are identified. We start with the simplest example of a free real scalar field. The action of a free (massless) scalar field $\Phi$ in 5D is
\begin{equation}
S = \int d^5 x\, \frac{1}{2} \del_M \Phi (x^\mu, y) \del^M \Phi(x^\mu, y).
\end{equation}
The scalar field has mass dimension 3/2 in 5D. Because the extra dimension is compactified, the field value is periodic in $y$ coordinates, $\Phi(x^\mu, y+2\pi R) = \Phi(x^\mu, y)$. We can perform the Fourier decomposition of $\Phi$ along the $y$ direction:
\begin{equation}
\Phi(x^\mu, y) = \frac{1}{\sqrt{2\pi R}} \sum_{n=-\infty}^{\infty} \phi^{(n)} (x^\mu) e^{i \frac{n}{R} y}.
\end{equation}
Reality of $\Phi$ implies $\phi^{(-n)} = {\phi^{(n)}}^\dagger$.
Plugging the Fourier series expansion into the action,
\begin{eqnarray}
S&=& \int d^4 x \,dy \;\frac{1}{2\pi R} \sum_{m,n} \Bigg[ \frac{1}{2} \del_\mu \phi^{(m)} (x) e^{i\frac{m}{R} y} \del^\mu \phi^{(n)} e^{i\frac{n}{R} y} \nonumber \\ 
&&   \qquad \qquad \qquad - \frac{1}{2} \bigg(i\frac{m}{R}\bigg) \phi^{(m)}(x) e^{i\frac{m}{R}y} \bigg(i\frac{n}{R}\bigg) \phi^{(n)}(x) e^{i\frac{n}{R}y} \Bigg] \nonumber \\
&=& \int d^4x \sum_{m,n} \Bigg( \int dy \; \frac{1}{2\pi R} e^{i \frac{m+n}{R} y}\Bigg) \frac{1}{2} \Bigg[ \del_\mu \phi^{(m)} (x) \del^\mu \phi^{(n)}(x) + \frac{mn}{R^2} \phi^{(m)}(x) \phi^{(n)}(x) \Bigg] \nonumber \\
&=& \int d^4 x  \; \frac{1}{2} \Bigg[ \sum_n \del_\mu \phi^{(-n)} \del^\mu \phi^{(n)} - \frac{n^2}{R^2} \phi^{(-n)} \phi^{(n)} \Bigg] \nonumber \\
&=& \int d^4 x \; \Bigg\{ \frac{1}{2} \del_\mu \phi^{(0)} \del^\mu \phi^{(0)} + \sum_{n=1}^{\infty} \bigg[ \del_\mu {\phi^{(n)}}^\dagger \del^\mu \phi^{(n)} - \frac{n^2}{R^2} {\phi^{(n)}}^\dagger \phi^{(n)} \bigg] \Bigg\}.
\end{eqnarray}
We can see from the 4D point of view that the action describes an (infinite) series of particles (Kaluza-Klein tower) with masses $m_{(n)} = n/R$. If the field $\Phi(x^\mu, y)$ has a 5D mass $m_0$, then the 4D Kaluza-Klein particles will have masses, $m_{(n)}^2 = m_0^2 + n^2/ R^2$. It is also easy to generalize it to higher dimensions compactified on a torus. In this case the masses of the Kaluza-Klein states are given by
\begin{equation}
m_{n_5,n_6,\ldots}^2 = m_0^2 + \frac{n_5^2}{R_5^2} + \frac{n_6^2}{R_6^2} + \cdots ,
\end{equation}
where $R_5,\,R_6, \ldots$ are the radii of the corresponding compact dimensions.

The next example is a gauge field in 5D, $A_M(x^\mu, y)$. We can similarly perform a Fourier decomposition along the compact dimension,
\begin{equation}
A_M(x^\mu, y) = \frac{1}{\sqrt{2\pi R}} \sum_n A_M^{(n)} (x^\mu) e^{i\frac{n}{R} y}.
\end{equation}
The derivative along the extra dimension can be replaced by $\del_5 \to i(n/R)$ under the Fourier decomposition. The action becomes
\begin{eqnarray}
S &=& \int d^4 x\, dy \bigg[-\frac{1}{4} F_{MN} F^{MN} \bigg] \nonumber \\
&=&  \int d^4 x \, dy  \bigg[ -\frac{1}{4} F_{\mu\nu} F^{\mu\nu} + \frac{1}{2} (\del_\mu A_5 - \del_5 A_\mu )(\del^\mu A_5 - \del_5 A^\mu) \bigg] \nonumber \\
&=& \int d^4 x \sum_n \Bigg[ \frac{1}{4} F_{\mu\nu}^{(-n)} F^{(n)\mu\nu} \nonumber \\
&& \qquad \qquad + \frac{1}{2} \bigg( \del_\mu A_5^{(-n)} + i \frac{n}{R} A_\mu^{(-n)}\bigg)\bigg( \del^\mu A_5^{(n)} - i\frac{n}{R} A^{(n)\mu}\bigg)\Bigg].
\end{eqnarray}
We can perform a gauge transformation to make $A_5$ constant along the extra dimension to remove the mixed terms,
\begin{eqnarray}
A_\mu^{(n)} &\to& A_\mu^{(n)} - i \frac{1}{n/R} \del_\mu A_5^{(n)},\nonumber \\
A_5^{(n)} &\to & 0, \qquad \qquad \qquad \qquad \mbox{ for } n\neq 0. 
\end{eqnarray}
In this gauge,
\begin{eqnarray}
S &=& \int d^4 x \Bigg\{ \bigg( -\frac{1}{4} F_{\mu\nu}^{(0)} F^{(0)\mu\nu} + \frac{1}{2} \del_\mu A_5^{(0)} \del^\mu A_5^{(0)} \bigg) \nonumber \\
&& \qquad \qquad + \sum_{n\geq 1} 2\bigg( -\frac{1}{4} F_{\mu\nu}^{(-n)} F^{(n)\mu\nu} + \frac{1}{2}\, \frac{n^2}{R^2} A_{\mu}^{(-n)} A_\mu^{(n)} \bigg) \Bigg\}.
\end{eqnarray}
We can see that zero modes contain a 4D gauge field and a real scalar. (For a non-Abelian gauge group the scalar would be in the adjoint representation.) For nonzero modes, $A_5^{(n)}$ is eaten and becomes the longitudinal mode of the corresponding massive vector field $A_\mu^{(n)}$. There is no scalar mode left for nonzero KK levels.

To match the 5D gauge coupling and the 4D gauge coupling we can examine the 5D covariant derivative $D_M = \del_M + i g_5 A_M$. Because $A_M$ has mass dimension 3/2 in 5D, the 5D gauge coupling $g_5$ has mass dimension $-1/2$. Expanding $A_\mu$ into its KK levels,
\begin{equation}
D_\mu = \del_\mu + i g_5 A_\mu = \del_\mu + i g_5 \frac{1}{\sqrt{2\pi R}} A_\mu^{(0)} + \cdots ,
\end{equation}
we see that the 4D gauge coupling is given by
\begin{equation}
g_4 = \frac{g_5}{\sqrt{2\pi R}}.
\end{equation}
Note that the 5D (also higher dimensional) gauge coupling has a negative mass dimension, so the 5D gauge theory is non-renormalizable. It becomes strongly interacting at energy scale $E \sim 1/g_5^2 = 1/(2\pi R g_4^2)$. Thus it can only be treated as a low energy effective theory with a cutoff $\Lambda \sim 4\pi/ g_5^2$. From the 4D point of view, the strong interaction comes from the enhancement of the number of KK modes accessible at the energy scale. The effective coupling is $\sqrt{N_{KK}} g_4$.

More generally, if we start with a $4+n$ dimensional gauge theory with $n$ dimensions compactified on a torus, the zero modes will contain a 4D gauge field together with $n$ adjoint scalars, and each nonzero KK level will have a 4D massive vector field and $(n-1)$ massive adjoint scalars.

Next we consider the gravitational field in $D= 4+n$ dimensions. Gravitational field is described by a symmetric metric tensor $g_{MN} =\eta_{MN} + h_{MN}$. There are $D(D+1)/2$ independent components of a symmetric tensor in $D$ dimensions. Many degrees of freedom can be removed by the $D$-dimensional general coordinate transformation $h_{MN} \to h_{MN}+\del_M \xi_N +\del_N \xi_M$. We can impose $D$ conditions to fix the gauge, {\it e.g.,} harmonic gauge, $ \del_M h_N^M = \frac{1}{2} \del_N h_M^M$. However, gauge transformations satisfying $\Box \xi_M=0$ are still allowed. Another $D$ conditions can be imposed. The number of independent degree of freedom  becomes
\begin{equation}
\frac{D(D+1)}{2} -2D = \frac{D(D-3)}{2}.
\end{equation}
For $D=4,5, 6$, they are listed below:\\
\begin{center}
\begin{tabular}{c|c}
Dimension($D$) & Number of degree of freedom \\
\hline
4 & 2 \\
5 & 5 \\
6 & 9 \\
\end{tabular}
\end{center}

On the other hand, a 4D massive spin-2 field has 5 polarizations. A 5D graviton with one spatial dimension compactified decomposes into
\begin{equation}
h_{MN} \supset h_{\mu\nu} \oplus h_{\mu 5} \oplus h_{55}.
\end{equation}
Zero modes consist of a 4D graviton, a massless vector, and a real scalar. For nonzero modes, $h_{\mu 5}^{(n)}$ and $h_{55}^{(n)}$ are eaten by $h_{\mu\nu}^{(n)}$ to form massive spin-2 fields.

The generalization to $4+n$ dimensions is straightforward. The zero modes contain one 4D graviton, $n$ massless vectors, and $n(n+1)/2$ scalars. For nonzero modes, each KK level has one massive spin-2 tensor, $(n-1)$ massive vectors, and $n(n+1)/2 -1 -(n-1) = n(n-1)/2$ massive scalars. One can also deduce the relation between the (reduced) Planck scales in 4 dimensions and higher dimensions,
\begin{eqnarray}
S &=& \frac{ \ov{M}_{4+n}^{2+n}}{2} \int d^{4+n}x \sqrt{|g_{(4+n)}|}\, R_{(4+n)} \nonumber \\
&=& \frac{ \ov{M}_{4+n}^{2+n}}{2} \, (2\pi R) ^n \int d^4 x \sqrt{-g_{(4)}}\, R_{(4)} + \cdots \nonumber \\
&=& \frac{\ov{M}_4^2}{2} \int d^4 x \sqrt{-g_{(4)}}\, R_{(4)} + \cdots \\ .
\end{eqnarray}
We obtain
\begin{equation}
\ov{M}_4^2 = (2\pi R)^n \ov{M}_{4+n}^{2+n} = V_n \ov{M}_{4+n}^{2+n} ,
\end{equation}
where the 4D reduced Planck scale 
\begin{equation}
\ov{M}_4 = \frac{1}{\sqrt{8\pi G_N}} = \frac{M_{pl (4)}}{\sqrt{8\pi}} = 2.4 \times 10^{18} \mbox { GeV}.
\end{equation}

The decomposition of a higher-dimensional graviton is summarized below~\cite{Giudice:1998ck,Han:1998sg}.

{\bf 4D graviton and its KK modes:} they are at the upper left $4\times 4$ corner of the $(4+n) \times (4+n)$ matrix for the higher-dimensional graviton,
\begin{equation}
\left(\begin{array}{ccc|ccccc} 
& G_{\mu\nu}^{\vec{k}} & & & & & & \\ 
\hline 
& & & & & & & \\ 
\end{array}
\right)
\end{equation}
These modes are labeled by the $n$-dimensional vector $\vec{k}$ which corresponds to the KK numbers along the various extra dimensions. 

 {\bf  4D vectors  and their  KK modes:} they live at the off-diagonal blocks of the higher-dimensional graviton matrix,
\begin{equation}
\left( \begin{array}{cccc|ccccc} 
& & & & & V_{\mu j}^{\vec{k}} & & \\ 
\hline 
& V_{\mu j}^{\vec{k}}& & & & & & \\ 
\end{array}
\right)
\end{equation}
For massive modes, they satisfy an additional constraint
\begin{equation}
\hat{k}^j V_{\mu j}^{\vec{k}} =0,
\end{equation}
so there are only $n-1$ independent vectors.

{\bf 4D scalars and their  KK modes:} the lower right $n \times n$ block of the graviton matrix corresponds to 
4D scalar fields: 
\begin{equation}
\left( \begin{array}{cccc|cccc} 
& & & & & & &\\ 
\hline 
& & & & & S^{\vec{k}}_{ij}& & \\ 
\end{array}
\right)
\end{equation}
The massive modes satisfy the additional constraint
\begin{equation}
\hat{k}^jS^{\vec{k}}_{jk}=0,
\end{equation}
It is also convenient to separate out the radion as a special field
which is represented by the trace $h^{\vec{k} j}_j$. The other scalars then satisfy the traceless condition
\begin{equation}
S^{\vec{k} j}_j =0.
\end{equation}

The explicit expressions for the canonically normalized 4D fields are given in unitary gauge by
(using the notation $\alpha=\sqrt{\frac{3(n-1)}{n+2}}$):
\begin{eqnarray}
&{\rm radion} & H^{\vec{k}}=\frac{1}{\alpha} h^{\vec{k} j}_j \nonumber \\
&{\rm scalars} & S^{\vec{k}}_{ij} = h^{\vec{k}}_{ij}-\frac{\alpha}{n-1} 
(\eta_{ij}+\frac{\hat{k}_i\hat{k}_j}{\hat{k}^2}) H^{\vec{k}} \nonumber \\
&{\rm vectors} & V_{\mu j}^{\vec{k}} =\frac{i}{\sqrt{2}}h^{\vec{k}}_{\mu j} \nonumber \\
&{\rm gravitons} & G_{\mu\nu}^{\vec{k}}=h^{\vec{k}}_{\mu \nu}+\frac{\alpha}{3}(\eta_{\mu\nu}
+\frac{\partial_\mu\partial_\nu}{\hat{k}^2}) H^{\vec{k}}.
\end{eqnarray}
The equation of motion in the presence of sources is given for the above fields by
\begin{equation}
(\Yfund +\hat{k}^2) \left( \begin{array}{c} G_{\mu\nu}^{\vec{k}} \\ V_{\mu j}^{\vec{k}} \\S^{\vec{k}}_{ij} \\
H^{\vec{k}} \end{array} \right) = \left( \begin{array}{c} 
\frac{1}{M_{Pl}}\left[-T_{\mu\nu}+(\eta_{\mu\nu}
+\frac{\partial_\mu\partial_\nu}{\hat{k}^2}) T^\mu_\mu/3 \right] \\ 0 \\ 0 \\
\frac{\alpha}{3 M_{Pl}} T^\mu_\mu \end{array} \right).
\end{equation}
We see that only the 4D graviton, the radion and their KK modes couple to the brane sources. Other fields do not couple to the matter fields on the brane directly and hence are not important for processes involving the brane matter fields.

\section{Effective Field Theory for a Three-brane}
\label{sec:eft}

In the large extra dimension scenario, the SM fields have to be localized on a 3-brane. To describe the interactions with the bulk, we need to develop an effective field theory for a 3-brane. We follow the work by Sundrum~\cite{Sundrum:1998sj}. A 3-brane breaks the higher-dimensional space-time symmetry either spontaneously or explicitly. In the case of spontaneous symmetry breaking, the 3-brane can fluctuate and there are Nambu-Goldstone bosons associated with the symmetry breaking. Examples are domain walls and D-branes in string theories.  In the case that the 3-brane does no fluctuate such as the orbifold fixed points or orientifolds in string theory, the extra space-time symmetry is explicitly broken and there is no associated Nambu-Goldstone mode. Here we will discuss the spontaneous symmetry breaking case.

We assume that the 3-brane is flat ($={\cal R}^4$) and the extra dimensions are compactified on a torus ${\cal T}^n$. Coordinates in the bulk are denoted as $X^M, \, M=0,1,2,\ldots, 3+n$, coordinates on the brane are denoted as $x^\mu, \, \mu =0, 1, 2, 3$, and the coordinates along the extra dimensions are $X^m, \, m=4, 5, \ldots, 3+n$. The metric in $4+n$ dimensions is $G_{MN}(X)$. The bulk coordinates describing the position occupied by a point $x$ on the 3-brane are denoted as $Y^M(x)$ and they are dynamical fields. SM fields are functions of $x$, $\phi(x),\, A_\mu(x),\, \psi(x)$.

The effective field theory describes small fluctuations around the vacuum state. In the vacuum
\begin{eqnarray}
G_{MN}(X) &=& \eta_{MN} \qquad (\mbox{using the mostly negative signature } +,-,-,\cdots), \nonumber \\
Y^M(x) &=& \delta_\mu^M x^\mu \qquad \mbox{(a simple gauge choice)}.
\end{eqnarray}
The bulk action is given by
\begin{equation}
S_{bulk} = \int d^{4+n} x \sqrt{|G|} \left( \frac{\ov{M}_{4+n}^2}{2} R_{(4+n)} - \Lambda \right).
\end{equation}
To write down an effective action for the brane localized fields we need the induced metric on the brane,
\begin{eqnarray}
ds^2 = G_{MN} dY^M(x) dY^N(x) &=& G_{MN} \frac{\del Y^M}{\del x^\mu } d x^\mu \frac{\del Y^N}{\del x^\nu} dx^\nu \nonumber \\ &=& g_{\mu\nu} d x^\mu d x^\nu .
\end{eqnarray}
We see that the induced metric is given by
\begin{equation}
g_{\mu\nu} = G_{MN} \frac{\del Y^M}{\del x^\mu }\frac{\del Y^N}{\del x^\nu}
\end{equation}
and in the vacuum state $g_{\mu\nu} = \eta_{\mu\nu}$.

The action is invariant under the general coordinate transformations of the bulk coordinates $X^M$ and also the general coordinate transformations of the brane coordinates $x^\mu$, so we need to contract the indices $M$ and $\mu$ separately. The brane action can be written as
\begin{eqnarray}
S_{brane} &=& \int d^4 x \sqrt{|g|} \Bigg\{ \frac{\tilde{M}_4^2}{2} R_{(4)} - f^4 + \frac{1}{2} g^{\mu\nu} D_\mu \phi D_\nu \phi - V(\phi) \nonumber \\
&& \qquad \qquad - \frac{g^{\mu\nu}g^{\rho\sigma}}{4} F_{\mu\rho} F_{\nu \sigma} + \cdots \Bigg\},
\end{eqnarray}
where $f^4$ is the brane tension, and the 4D Planck scale receives contributions from both the bulk term $\ov{M}_{4+n}^{2+n} V_n$ and the brane term $ \tilde{M}_4^2$. Assuming that the brane tension $f \ll \ov{M}_{4+n}$, then we can ignore the back reaction on gravity.

We can use the 4D reparametrization invariance to gauge fix
\begin{equation}
Y^\mu (x) = x^\mu, \quad \mu = 0, 1, 2, 3,
\end{equation}
then only $Y^m(x), \, m= 4, 5, \cdots n+3$ are physical degrees of freedom. Their kinetic terms can be derived by expanding the tension term of the brane action,
\begin{eqnarray}
S &=& \int d^4 x \sqrt{|g|} [-f^4 + \cdots],  \nonumber \\
&& g_{\mu\nu} = G_{MN} \del_\mu Y^M \del_\nu Y^N = \eta_{\mu\nu} + \del_\mu Y^m \del_\nu Y_m + \cdots , \nonumber \\
&& \det g = -1 - \del_\mu Y^m \del ^\mu Y_m + \cdots , \nonumber \\
&& \sqrt{|g|} = 1 + \frac{1}{2}  \del_\mu Y^m \del^\mu Y_m + \cdots , \nonumber \\
S &=& \int d^4 x (-f^4) \left( 1 + \frac{1}{2} \del_\mu Y^m \del^mu Y_m + \cdots\right) \nonumber \\
&=& \int d^4 x \left[ (-f^4) + \frac{f^4}{2} \del_\mu Y_m \del^\mu Y_m + \cdots \right] \quad ( Y^m= -Y_m) .
\end{eqnarray}
Canonically normalized fields are given by $Z_m = f^2 Y_m$. A positive tension $(f^4>0)$ implies a positive kinetic term. On the other hand, if the brane tension is negative, then the kinetic term is negative and $Y_m$'s are ghost field. The system is unstable and the brane would like to crumble. Negative tension can therefore only occur with explicit breaking of higher-dimensional space-time symmetry where there is no Nambu-Goldstone bosons $Y_m$.

We can also derive the couplings of the localized SM fields to bulk gravitons. The stress-energy tensor of the SM fields is given by
\begin{equation}
T^{\mu\nu} = \frac{2}{\sqrt{|g|}} \frac{\delta S}{\delta g_{\mu\nu}},
\end{equation}
and the interaction is 
\begin{equation}
S_{\rm int} \supset \int d^4 x \sqrt{|g|} \,\frac{1}{2}\, T^{\mu\nu} \delta g_{\mu\nu} .
\end{equation}
Expanding $g_{\mu\nu}$ we have
\begin{eqnarray}
g_{\mu\nu} &=& G_{MN} \del_\mu Y^M \del_\nu Y^N \nonumber \\
&=& ( \eta_{MN} + \kappa_{4+n}  H_{MN}) ( \delta_\mu^M + \cdots) (\delta_\nu^N + \cdots) \nonumber \\
&=& \eta_{\mu\nu} + \frac{2}{\ov{M}_{4+n}} H_{\mu\nu} + \cdots .
\end{eqnarray}
For extra dimensions compactified on a torus, $H_{MN}(X^L)$ is periodic for $L=4,5,\cdots 3+n$,
\begin{equation}
H_{MN} (x, y) = \sum_{k_1 = -\infty}^{\infty} \cdots \sum_{k_n = -\infty}^{\infty} \frac{h_{MN}^{(\vec{k})}}{\sqrt{V_n}} e^{i \frac{\vec{k} \cdot \vec{y}}{R} }.
\end{equation}
We can choose $\vec{y}=0$ for the 3-brane location, then
\begin{equation}
H_{MN} (x, 0) = \sum_{\vec{k}} \frac{h_{MN}^{(\vec{k})}}{\sqrt{V_n}}.
\end{equation}
The interactions between the SM fields and KK gravitons are given by
\begin{eqnarray}
S_{int} &\supset & \int d^4 x \sqrt{|g|}\, \frac{1}{2}\, T^{\mu\nu} \sum_{\vec{k}} \frac{2}{\ov{M}_{4+n} \sqrt{V_n} } h_{\mu\nu}^{(\vec{k})} \nonumber \\
&=& \int d^4 x \sqrt{|g|}\, \frac{1}{\ov{M_4}}\, T^{\mu\nu} \sum_{\vec{k}} h_{\mu\nu}^{(\vec{k})} .
\end{eqnarray}

\section{Experimental Constraints and Tests of Large Extra Dimensions}
\label{sec:exp}

Large extra dimensions may appear to be a radical proposal to address the hierarchy problem. Whenever some possible new physics is proposed, we need to ask whether it is consistent with the current experimental constraints and how we can test it experimentally. Some characteristics of large extra dimensions are crucial for experimental constraints and tests:
\begin{itemize}
\item TeV cutoff: Precision electroweak tests and high energy collisions are sensitive to higher-dimensional operators suppressed by the TeV scale.

\item Light degrees of freedom: They can appear as missing energies at colliders and rare decays of unstable particles. They also affect astrophysics ({\it e.g.,} star cooling) and cosmology ({\it e.g.,} expansion rate of the universe).

\item Long-lived KK gravitons: They can affect astrophysics (diffuse $\gamma$-ray background from late decays of long-lved particles) and cosmology (over-closure of the universe).
\end{itemize}
We will discuss various experimental bounds and possible tests in the following subsections. We will only give a flavor of what types of constraints are relevant and make simple estimates. For more detailed discussion please consult the original papers in the literature.

\subsection{Laboratory bounds on long-range forces}
\begin{figure}
\begin{center}
\epsfig{file=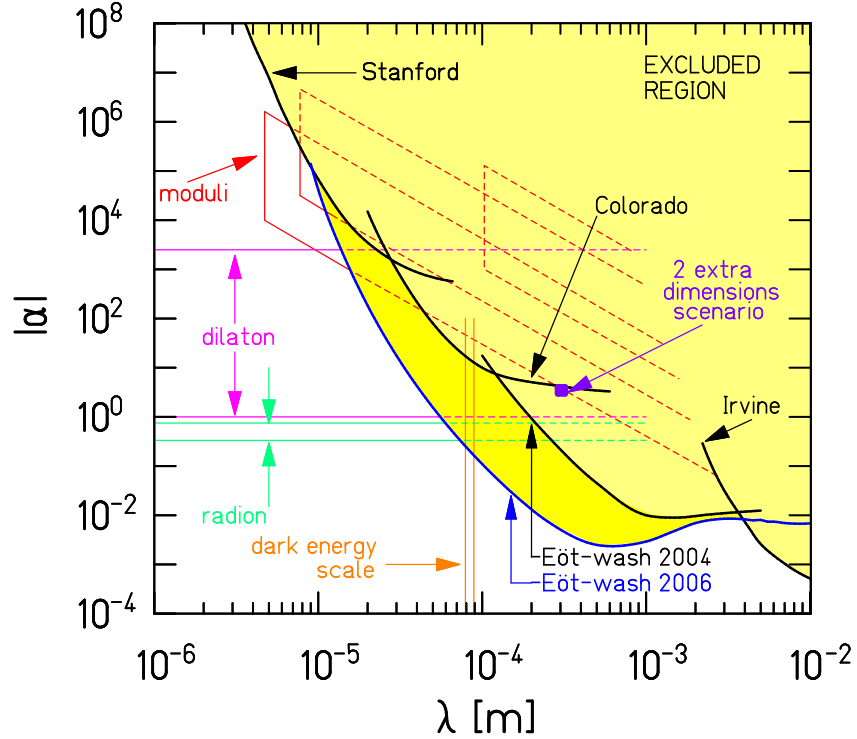,width=5.0in}
\end{center}
\caption{Laboratory bounds on deviations of the gravitational inverse-square law, taken from D.~J.~Kapner {\it et al}~\cite{Kapner:2006si}.}
\label{fig:lab_bounds}
\end{figure}

At distances shorter than the compactification radius of extra dimensions, the gravitational force will be modified from the usual inverse-square law. The experimental tests of the gravitational force are usually parametrized by the modified potential,
\begin{equation}
V(r) = - G_N \frac{m_1 m_2}{r} \left( 1 + \alpha e^{-r/\lambda}\right),
\end{equation}
where $\lambda$ is the distance where the modification occurs and is given by the inverse mass of the new light particle which mediates the new force, and $\alpha$ represents the strength of the new force relative to the gravitational force. For the large extra dimension scenario, $\lambda$ is the inverse mass of the first KK graviton, $\lambda = (m^{(1)})^{-1} = R$. and $\alpha$ is the number of the first KK modes ({\it e.g.,} $\alpha=4$ for 2 extra dimensions on a torus). The bounds from various experiments are shown in Fig.~\ref{fig:lab_bounds}.
For 2 extra dimensions of the same size, the current bound is $R< 37 \mu$m~\cite{Kapner:2006si}. Using the relation between the reduced Planck scales in 4 dimensions and in 6 dimensions,
\begin{equation}
\ov{M}_4^2 = (2\pi R)^2 \ov{M}_6^4,
\end{equation}
we obtain a bound on the 6-dimensional Planck scale,
\begin{equation}
\ov{M}_6 = \sqrt{\frac{\ov{M}_4}{2\pi R}} > 1.4 \mbox{ TeV}.
\end{equation}
As we will see later, there are more stringent bounds from astrophysics and cosmology. For more than 2 extra dimensions, the laboratory constraints on long-range forces do not place any significant bounds on the higher-dimensional Planck scale.

\subsection{Particle physics constraints}
The constraints on large extra dimensions from particle physics can be divided into several categories.

\subsubsection{Higher dimensional operators}
If the fundamental Planck scale as a cutoff of the SM physics is at $\sim$TeV, we expect that there will be higher-dimensional operators suppressed by the cutoff scale,
\begin{equation}
{\cal L} \supset \frac{c_i}{\Lambda^m} {\cal O}_i^{4+m} \mbox{ with } \Lambda \sim M_{4+n} \sim \mbox{ TeV}.
\end{equation}
\begin{enumerate}
\item Some operators are strongly constrained, {\it, e.g.,}
\begin{eqnarray}
\Lambda_{\rm proton\, decay} &\gtrsim & 10^{15} \mbox{ GeV}, \nonumber \\
\Lambda_{\rm Majorana\, neutrino} &\sim& 10^{15} \mbox{ GeV}, \nonumber \\
\Lambda_{\rm FCNC, \; \not \!\!CP} &\gtrsim& 100 \sim 1000 \mbox{ TeV}.
\end{eqnarray}
They appear to be dangerous (but one faces the same problem in any extension of the SM at the TeV scale). However, these operators violate (approximate) symmetries of SM, so one can imagine ways to suppress them, such as gauging baryon number symmetry and imposing flavor symmetries etc.

\item For operators which respect symmetries of SM, the constraints come from the electroweak precision tests. In general $\Lambda \gtrsim $ hundreds GeV $\sim$ 10 TeV. They are reviewed in Prof.~Skiba's lectures~\cite{Skiba}. It causes a little hierarchy problem if $M_{4+n} \sim $ 1 TeV.

\item Some operators can be induced by KK graviton exchanges~\cite{Giudice:1998ck,Han:1998sg,Nussinov:1998jt,Hewett:1998sn}. For example $e^+ e^- \to e^+ e^-$ through KK gravitons (Fig.~\ref{fig:eeee}) induces an operator,
\begin{figure}
\begin{center}
\epsfig{file=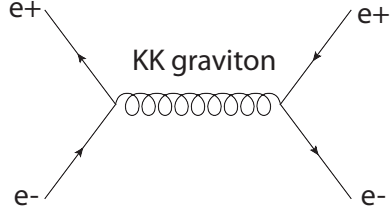,width=2.0in}
\end{center}
\caption{A four fermion interaction induced by KK graviton exchanges.}
\label{fig:eeee}
\end{figure}
\begin{equation}
{\cal L}_{\cal O} \sim c \sum_{\vec{k}} \frac{E^2}{\ov{M}_4^2} \frac{1}{| \vec{k} R^{-1}|^2} \left( \ov{\psi} \psi \right)^2,
\end{equation}
where we have omitted the tensor structure.
The sum is logarithmic divergent for $n=2$ and power divergent for $n>2$, so it needs to be cut off,
\begin{equation}
{\cal L}_{\cal O} \simeq \frac{E^2 \Lambda^{n-2}}{\ov{M}_{4+n}^{2+n}} \left( \ov{\psi} \psi \right)^2.
\end{equation}
If we take the cutoff $\Lambda \sim \ov{M}_{4+n}$, then $\ov{M}_{4+n} \gtrsim$ 1 TeV from current experimental constraints. Another example is muon $g-2$ where the KK graviton loop gives rise to a divergent contribution which needs to be cut off (Fig.~\ref{fig:muon_g-2})~\cite{Graesser:1999yg},
\begin{figure}
\begin{center}
\epsfig{file=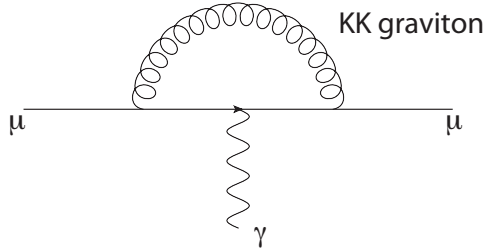,width=2.5in}
\end{center}
\caption{A Feynman diagram which contribute to muon $g-2$ with KK graviton loop.}
\label{fig:muon_g-2}
\end{figure}
It gives similar constraint.
\end{enumerate}

\subsubsection{Real emission of KK gravitons}
This set of experimental tests relies on the existence of light KK gravitons. It is less sensitive to the cutoff physics and hence more robust. The mass splitting of KK modes in large extra dimensions is
\begin{equation}
\Delta m \sim \frac{1}{R} \sim \ov{M}_{4+n} \left( \frac{\ov{M}_{4+n}}{\ov{M}_4}\right)^{\frac{2}{n}} = \left(\frac{\ov{M}_{4+n}}{\rm TeV}\right)^{\frac{n+2}{n}} 10^{\frac{12n-31}{n}} \mbox{ eV}.
\end{equation}
It is extremely small in particle physics scale and the KK spectrum can be treated as continuum for any practical purpose. We can estimate the number of KK modes with momentum in extra dimensions between $k$ and $k+dk$,
\begin{eqnarray}
dN &=& S_{n-1} k^{n-1} dk \nonumber \\
&=& S_{n-1} m^{n-1} R^{n-1} dm \cdot R \nonumber \\
&=& \frac{S_{n-1}}{(2\pi)^n} \frac{\ov{M}_4^2}{\ov{M}_{4+n}^{n+2}} m^{-1} dm ,
\end{eqnarray}
where $S_{n-1} = (2\pi)^{\frac{n-1}{2}}/ \Gamma(\frac{n-1}{2})$ is the surface area of an $n-1$ dimensional sphere with unit radius, and $m = |\vec{k}|/R$. The number of KK modes accessible at energy $E$ is 
\begin{equation}
N(E) \sim \frac{\ov{M}_4^2}{\ov{M}_{4+n}^2} \left( \frac{E}{\ov{M}_{4+n}}\right)^n.
\end{equation}
We see that there are more accessible KK modes at higher energies, which implies stronger constraints will come from processes with higher energies. On the other hand, for $E< \ov{M}_{4+n}$, there are fewer KK modes for larger $n$. It implies weaker constraints for larger $n$ from the same process. 

\begin{enumerate}
\item Rare decays to KK gravitons: An example is $K \to \pi +$ KK graviton (Fig.~\ref{fig:rare_decay})~\cite{ArkaniHamed:1998nn}.
\begin{figure}
\begin{center}
\epsfig{file=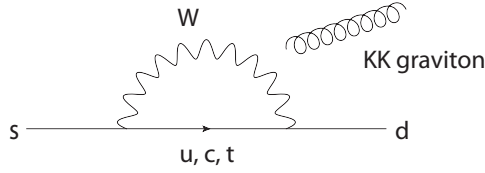,width=2.5in}
\end{center}
\caption{A diagram which contribute to $K \to \pi +$ KK graviton decay.}
\label{fig:rare_decay}
\end{figure}
In the 4D picture, decay width to a single KK graviton is 
\begin{equation}
\Gamma_1 \sim \frac{1}{16\pi} \frac{m_K^5}{m_W^4}\frac{m_K^2}{\ov{M}_4^2}.
\end{equation}
The number of KK gravitons with mass less than $m_K$ is $\sim (m_K R)^n$, so the total decay width of $K$ to $\pi$ + KK graviton is
\begin{eqnarray}
\Gamma_{K\to \pi + {\rm KK\, graviton}} &\sim&   \frac{1}{16\pi} \frac{m_K^5}{m_W^4}\frac{m_K^2}{\ov{M}_4^2} \times m_K^n R^n \nonumber \\
& \sim & \frac{1}{16\pi} \frac{m_K^5}{m_W^4}\left(\frac{m_K}{\ov{M}_{4+n}}\right)^{n+2}.
\end{eqnarray}
We see that the result is proportional to the square of the higher dimensional graviton coupling $ \ov{M}_{4+n}^{-\frac{n+2}{2}}$ as expected. One can obtain the same estimate from a higher dimensional picture. The current experimental bound on the rare $K$ decay is
\begin{equation}
B(K\to \pi + X) < 10^{-10}, \; \tau(K) \sim 10^{-8} {\rm s} \Rightarrow \Gamma(K\to \pi +X) < 10^{-26} \mbox{ GeV}.
\end{equation}
For $n=2$, we have $\ov{M}_6 \gtrsim 1$ TeV and the constraints are quite weak for $n>2$.

\item Production of KK gravitons at high energy colliders: KK gravitons couple to the stress-energy tensor $T_{\mu\nu}$, so they can be attached anywhere in a process. The leading processes for KK graviton production in high energy collisions are
\begin{eqnarray}
e^+ e^- & \to& \gamma/Z + G_{KK} \nonumber \\
q \bar{q} &\to& g + G_{KK} \nonumber \\
q g &\to & q+G_{KK}
\end{eqnarray}
The Feynman diagrams for these processes are shown in Fig.~\ref{fig:feynman}, and the Feynman rules can be found in Ref.~\refcite{Giudice:1998ck,Han:1998sg}. 
\begin{figure}
\begin{center}
\epsfig{file=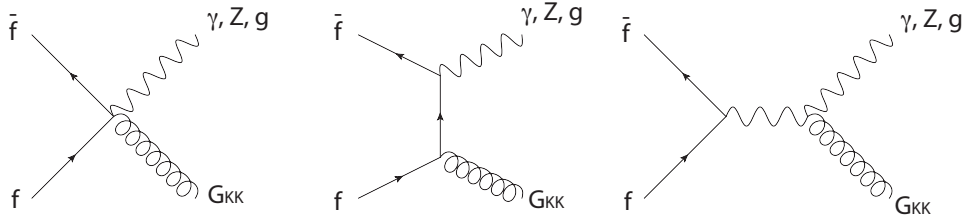,width=5.0in}
\end{center}
\caption{Feynman diagrams for the KK graviton production.}
\label{fig:feynman}
\end{figure}
The experimental signals are missing energy/momentum from the unobserved graviton~\cite{Giudice:1998ck,Han:1998sg,Mirabelli:1998rt}. The current bounds can be found in Ref.~\refcite{Landsberg:2008ax} which is reproduced here. 
%
\begin{center}
\begin{tabular}{lccccccc}
\hline
Experiment and channel 	& $n=2$ & $n=3$ & $n=4$ & $n=5$ & $n=6$ \\
\hline
LEP Combined & 1.60 & 1.20 & 0.94 & 0.77 & 0.66 \\
\hline
CDF monophotons, 2.0 fb$^{-1}$  &   1.08    &  1.00   & 0.97 &  0.93  & 0.90  \\
D\O\ monophotons, 2.7 fb$^{-1}$   &   0.97    &  0.90   & 0.87 &  0.85  & 0.83 \\
\hline
CDF monojets, 1.1 fb$^{-1}$  & 1.31  & 1.08 & 0.98 & 0.91 & 0.88 \\
\hline
CDF combined & 1.42 & 1.16 & 1.06 & 0.99 & 0.95 \\
\hline
\end{tabular}
\end{center}
%

\end{enumerate}

\subsubsection{Black hole productions at colliders and in cosmic rays}
For $M_{pl,4+n} \sim$ TeV and $E_{\rm CM} >M_{pl,4+n}$, black holes can be produced at high energy collisions~\cite{Dimopoulos:2001hw,Giddings:2001bu}.
Semiclassical arguments work for $M_{BH} \gg M_{pl,4+n}$.

The Schwarzschild radius for a $4+n$-dimensional black hole is
\begin{equation}
R_s \sim \frac{1}{M_{pl,4+n} }\left( \frac{M_{BH}}{M_{pl,4+n}}\right)^{\frac{1}{n+1}} .
\end{equation}
For 2 partons with $\sqrt{s}> M_{pl,4+n}$ moving in opposite directions, a black hole forms with mass $M_{BH} \sim \sqrt{s}$ if the impact parameter is smaller than $R_s$. The cross section is given by the geometrical formula,
\begin{equation}
\sigma( M_{BH}) \approx \pi R_s^2 \sim \frac{1}{M_{pl,4+n}^2} \left( \frac{M_{BH}}{M_{pl,4+n}}\right)^{\frac{2}{n+2}},
\end{equation}  
which is large for $\sqrt{s} \gg M_{pl,4+n}$.

The produced black holes will decay through Hawking radiation with the Hawking temperature
\begin{equation}
T_H \sim M_{pl,4+n} \left( \frac{M_{pl,4+n}}{M_{BH}}\right)^{\frac{1}{n+1}} \sim \frac{1}{R_s} .
\end{equation}
A black hole decays equally to a particle on the brane or in the bulk, so it decays mostly to the brane if there are more brane (SM) fields than the bulk fields (graviton). The multiplicity of particles produced in a black hole evaporation can be estimated as
\begin{equation}
\langle N \rangle = \left\langle \frac{M_{BH}}{E_{\rm particle}} \right\rangle \sim \left\langle \frac{M_{BH}}{T_{H}} \right\rangle \sim \left( \frac{M_{BH}}{M_{pl,4+n}}\right)^{\frac{n+2}{n+1}} .
\end{equation}
The branching fraction to leptons is about 10\% and to photons is about 2\%. The search strategy is to select events with high multiplicities and $e^{\pm}$ or $\gamma$ with $E> 100$ GeV. The reach at the LHC extends up to $M_{pl,4+n} \sim 9$ TeV~\cite{Landsberg:2008ax}. High energy cosmic rays also provide powerful probes of black hole productions in the large extra dimension scenario~\cite{Feng:2001ib,Anchordoqui:2001ei,Emparan:2001kf,Anchordoqui:2001cg}.

\subsection{Astrophysics bounds}

Astrophysics provides some of the strongest constraints on the large extra dimension scenario (for small $n$).

\begin{enumerate}
\item Star cooling: Constraints on the fundamental Planck scale can be obtained from bounds on energy loss due to KK graviton emission from the Sun, red giants, and supernovae. The rate for KK graviton emission can be estimated as
\begin{equation}
\Gamma \sim \frac{1}{\ov{M}_4^2} (T R)^n \sim \frac{T^n}{\ov{M}_{4+n}^{n+2}}.
\end{equation}
We see that stronger bounds are obtained for higher temperature, and smaller $n$. The temperatures for various stars are
\begin{center}
\begin{tabular}{ll}
Sun & $T\sim $ keV , \\
Red Giants & $T \sim $ 100 keV ,\\
SN1987A & $T \sim$ 30 MeV .
\end{tabular}
\end{center}

We will discuss the bounds from the supernova SN1987A as an example. The energy loss of SN1987A is mostly due to neutrino emission. We can obtain a quick estimate from the bound on the axion. The axion coupling to nucleons is $\sim m_N/ f$ where $f$ is the axion decay constant. On the other hand, the neutrino coupling is $\sim T_\nu^2/ M_W^2$. Neutrinos are emitted from a neutrino sphere at $R\sim $ 15 km with $\rho \sim 10^{12}$ g/cm$^3$, $T_\nu \sim $ 4 MeV, beyond which there is no more scattering. The effective neutrino coupling is 
\begin{equation}
\frac{T_\nu^2}{M_W^2} \sim 10^{-9}.
\end{equation}
Requiring the axion coupling to be smaller than the neutrino coupling we obtain
\begin{equation}
\frac{m_N}{f} < 10^{-9} \quad \Rightarrow \quad f\gtrsim 10^9 \mbox{ GeV}.
\end{equation}
Now comparing the KK graviton production cross section $\sim T^n/ \ov{M}_{4+n}^{2+n}$ and the axion production cross section $\sim 1/f^2$, we obtained the following dictionary for converting experimental bounds,
\begin{equation}
\frac{1}{f^2} \leftrightarrow \frac{T^n}{\ov{M}_{4n}^{2+n}} \quad \Rightarrow \quad \ov{M}_{4+n} \gtrsim ( T^n f_{\rm lower\, bound}^2)^{\frac{1}{n+2}}.
\end{equation}
For $T \sim 30$ MeV, we have 
\begin{eqnarray}
\ov{M}_{4+n} &\gtrsim& 5 \mbox{ TeV} , \qquad n=2, \nonumber \\
&\gtrsim& 500 \mbox{ GeV},\quad  n=3.
\end{eqnarray}
More precise calculations lead to~\cite{Cullen:1999hc,Barger:1999jf,Hanhart:2001fx} 
\begin{eqnarray}
M_6 &\gtrsim& 14 \mbox{ TeV}, \nonumber \\
M_7 & \gtrsim & 1.6 \mbox{ TeV},
\end{eqnarray}
where $M_{4+n} = (2\pi)^{\frac{n}{2+n}} \ov{M}_{4+n}$.

\item Diffuse $\gamma$ ray background from long-lived KK graviton decays: If the KK gravitons decay back to photons, it will affect the diffuse $\gamma$ ray spectrum. The measurement of EGRET satellite put bounds on the higher dimensional Planck scale $M_6 > 38$ TeV, $M_7 > 4.1$ TeV~\cite{Hall:1999mk,Hannestad:2001jv}.

\item Most of the KK gravitons emitted by supernova remnants and neutron stars are gravitationally trapped. The gravitons forming this halo occasionally decay, emitting photons. Limits on $\gamma$ rays from neutron star sources imply $M_6 > 200$ TeV, $M_7 > 16$ TeV~\cite{Hannestad:2001xi}. The decay products of the gravitons forming the halo can hit the surface of the neutron star, providing a heat source. The low measured luminosities of some pulsars implies $M_6 > 750$ TeV, $M_7 > 35$ TeV~\cite{Hannestad:2001xi}.

\end{enumerate}

The last 2 constraints assume how gravitons decay and hence have some model-dependence. They may be evaded by some mild modifications of  the theory ({\it e.g.,} KK gravitons may decay into hidden stuffs in some other branes).

\subsection{Cosmological constraints}

The earliest time we know about cosmology with some certainty is the era of Big-Bang Nucleosynthesis (BBN) which starts at about $T\sim $ 1 MeV. In order to have BBN, the reheating temperature after inflation must be higher than 1 MeV. On the other hand, high reheating temperature means more KK gravitons are produced. They can affect
\begin{itemize}
\item cooling of the universe due to graviton emission into extra dimensions,
\item expansion of the universe during BBN as KK gravitons redshift as $R^{-3}$ (radiation domination is required during BBN),
\item over-closure of the universe if graviton energy density is too large.
\end{itemize}
They put constraints on the reheating temperature as well as the higher-dimensional Planck scale.

\section{Physics in the Bulk}

Large extra dimensions are an interesting proposal to address the hierarchy problem by bringing down the fundamental Planck scale to $\sim$ TeV. However, this also removed the possibility of using energy scales between the electroweak scale and the usual 4D Planck scale to address other questions in particle physics, such as the fermion mass hierarchies. Indeed, one of the major activities in particle physics is to explain small numbers ({\it e.g.,} fermion masses, smallness or absence of flavor-changing effects, proton decay, \ldots). The traditional approach is to employ symmetries. Extra dimensions provide some new ways to realize small numbers.

\begin{enumerate}
\item Large volume suppression: In addition to explain $(M_{4+n}/M_4)^2 = (M_{4+n}^n V_n)^{-1} \ll 1$, the large volume in extra dimensions can be used to explain other small couplings. For example, if right-handed neutrinos live in extra dimensions, their couplings to left-handed neutrinos localized on the brane are suppressed by the large volume factor just as the case for the gravitons. It can explain the smallness of the neutrino masses~\cite{Dienes:1998sb,ArkaniHamed:1998vp}.

\item Locality: Fields localized at different places in extra dimensions cannot couple directly. Imaging that a coupling of SM fields is forbidden by some symmetry and the symmetry is broken at a place far away from the SM in the extra dimensions. The symmetry breaking effect is mediated to the SM sector by some mediator field $\chi$. Then, the induced coupling of the SM fields can be highly suppressed~\cite{ArkaniHamed:1998sj}:
\begin{eqnarray}
\mbox{Suppression factor } & \sim& \frac{1}{r^{n-2}}, \;\;\; \mbox{for } r m_{\chi} \ll 1  \mbox{ (massless mediator),} \nonumber \\
&\sim & \frac{e^{-m_{\chi} r}}{r^{n-2}}, \; \mbox{for } r m_{\chi} \gg 1  \mbox{ (massive mediator),}
\end{eqnarray}
where $r$ is the distance between the SM brane and the source of the symmetry breaking in extra dimensions.
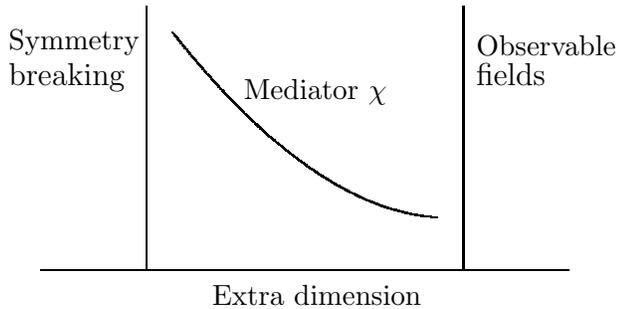
\begin{figure}
\begin{center}
\begin{picture}(240,115)(-120,5)
\put(-100,20){\line(1,0){200}}
\put(-60,20){\line(0,1){100}}
\put(60,20){\line(0,1){100}}
\qbezier(-50,110)(3,43)(50,40)
\put(-35,7){\small Extra dimension}
\put(-112,90){\shortstack[l]{\small Symmetry\\breaking}}
\put(65,90){\shortstack[l]{\small Observable\\fields}}
\put(-23,85){\small Mediator $\chi$}
\end{picture}
\end{center}
\caption{Suppression of symmetry breaking effects through extra dimensions.}
\label{fig:shining}
\end{figure}
\end{enumerate}

\section{Standard Model Fields in Extra Dimensions}

So far we have considered gravity-only extra dimensions while the SM fields live on a 3-brane. It is interesting to study the case that some of the SM fields also propagate in extra dimensions ({\it e.g.,} the 3-brane can have nonzero thickness). Because Standard Model has been tested up to the TeV scale, the size of such extra dimensions must be $\lesssim$ TeV$^{-1}$. Since there may or may not be other large extra dimensions where only gravity propagates, in discussing SM fields propagating in extra dimensions we will take the fundamental Planck scale as a free parameter and focus on the SM sector.

As SM fermions and Higgs carries gauge quantum numbers, they cannot propagate in extra dimensions unless the corresponding gauge fields also propagate in extra dimensions. On the other hand, SM fermions and Higgs may still be localized in 4 dimensions even if gauge fields propagate in extra dimensions. As a result, there are several possibilities:
\begin{enumerate}
\item SM gauge fields propagate in extra dimensions while fermions and Higgs live on a 3-brane: 
In this case, higher-dimensional operators are induced by KK gauge boson exchanges and they are strongly constrained by the electroweak precision data. For example, Fig.~\ref{fig:4fermion} shows a 4-fermion operator induced by the KK photons and $Z$-bosons. For one extra dimension, the size of the extra dimension is constrained to be  $R^{-1} > 6.6$ TeV~\cite{Cheung:2001mq}.
\begin{figure}
\begin{center}
\epsfig{file=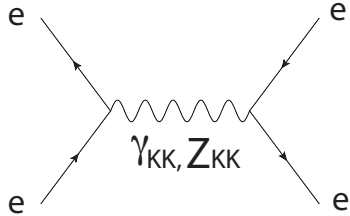,width=1.8in}
\end{center}
\caption{A four fermion interaction induced by KK gauge boson exchange.}
\label{fig:4fermion}
\end{figure}

Such a setup is useful for gaugino mediated SUSY breaking. If the SUSY is broken at a place away from the SM fermions and their superpartners in extra dimensions (Fig.~\ref{fig:gmsb}), SUSY breaking is only transmitted to the MSSM sector by the gauge sector which is flavor universal~\cite{Mirabelli:1997aj,Kaplan:1999ac,Chacko:1999mi}. Unwanted flavor-non-universal interactions with SUSY breaking sector are forbidden by locality in extra dimensions.  
\begin{figure}
\begin{center}
\begin{picture}(240,120)(-120,-60)
\put(-70,0){
\put(-20,-55){\line(0,1){90}}
\put(-20,-55){\line(2,1){40}}
\put(20,55){\line(0,-1){90}}
\put(20,55){\line(-2,-1){40}}
\put(-15,-8){\shortstack{\footnotesize MSSM\\matter}}
}
\put(70,0){
\put(-20,-55){\line(0,1){90}}
\put(-20,-55){\line(2,1){40}}
\put(20,55){\line(0,-1){90}}
\put(20,55){\line(-2,-1){40}}
\put(-18,-8){\shortstack{\footnotesize SUSY\\breaking}}
}
\put(-22,-8){\shortstack{\footnotesize MSSM\\gauge fields}}
\end{picture}
\end{center}
\caption{Gaugino mediated SUSY breaking through extra dimensions.}
\label{fig:gmsb}
\end{figure}
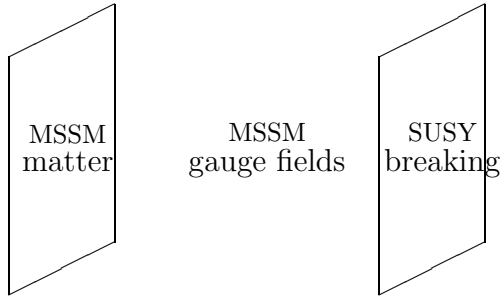

\item All SM fields live in the (same) extra dimensions: In this case they are called Universal Extra Dimensions (UEDs)~\cite{Appelquist:2000nn}. The experimental bound is weaker due to approximate KK number conservation, $R^{-1} \gtrsim 300-600$ GeV~\cite{Appelquist:2002wb,Gogoladze:2006br}.

\item It is also possible to have the mixed scenario. Different SM fermions may even be localized at different locations in extra dimensions. It can be used to explain the fermion mass hierarchy as we will see next.

\end{enumerate}

As mentioned earlier, gauge theories in more than 4 dimensions are non-renormalizable and should be treated as low energy effective theories below some cutoff $\Lambda$. The effective expansion parameter is $\frac{N_{KK} N_c \alpha}{4\pi}$. The theory becomes strongly coupled when $\frac{N_{KK} N_c \alpha}{4\pi}\sim 1$ or $N_{KK}\sim 40$, so one requires $\Lambda R \lesssim 40 $ for 1 extra dimension and $\Lambda R \lesssim 5$ for 2 extra dimensons. In the following subsections we discuss several interesting topics with SM living in extra dimensions.

\subsection{Split fermions -- fermion mass hierarchies without symmetries from extra dimensions}

Extra dimensions provide a new tool ``locality'' for doing physics in the bulk. In particular, we can use it to explain the absence or smallness of certain interactions. For example, the smallness of fermion Yukawa couplings and the absence of proton decays may be explained if SM fermions are localized at different places in extra dimensions~\cite{ArkaniHamed:1999dc}. To see that, we first consider how to localize a fermion in extra dimensions.

\begin{figure}
\begin{center}
\epsfig{file=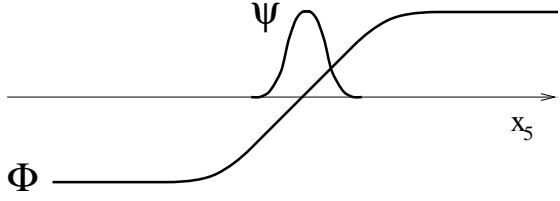,width=4.0in}
\end{center}
\caption{A chiral fermion localized by a domain wall.}
\label{fig:domain_wall}
\end{figure}

A chiral fermion can be localized in an extra dimension by a domain wall as shown in Fig.~\ref{fig:domain_wall}.
A fermion in 5 dimensions has 4 components. Its action is given by
\begin{equation}
S= \int d^4 x dy \ov{\Psi} \left[ i \Gamma^\mu \del_\mu + i \Gamma^5 \del_5 + \Phi(y)\right] \Psi ,
\end{equation}
and the corresponding Dirac equation is 
\begin{equation}
\left[ i \Gamma^\mu \del_\mu + i \Gamma^5 \del_5 + \Phi(y)\right] \Psi =0.
\end{equation}
In the Weyl basis,
\begin{equation}
\Gamma^\mu = \gamma^\mu = \begin{pmatrix} 0 & \sigma^\mu \\ \bar{\sigma}^\mu & 0 \end{pmatrix}, \; \Gamma^5 = i\begin{pmatrix}-I & 0 \\ 0& I\end{pmatrix}, \; \Psi = \begin{pmatrix}\psi_L \\ \psi_R\end{pmatrix},
\end{equation}
where $\sigma^\mu = (I, \vec{\sigma}), \; \bar{\sigma}^\mu = (I, -\vec{\sigma})$. There are two Lorentz invariant fermion bilinears, 
\begin{equation}
\ov{\Psi}_1 \Psi_2,\; \Psi_1^T C_5 \Psi_2, \mbox{ with } C_5 = \Gamma^0 \Gamma^2 \Gamma^5 = \begin{pmatrix}\epsilon & 0 \\ 0 & -\epsilon \end{pmatrix}.
\end{equation}
We look for solutions which are left- or right-handed 4D modes, $i \Gamma^5 \psi_L = \psi_L$, $i \Gamma^5 \psi_R = - \psi_R$,
\begin{equation}
\Psi(x, y) = \sum_n f_L^{(n)} (y) \psi_L^{(n)} (x) + \sum_n f_R^{(n)} (y) \psi_R^{(n)} (x) ,
\end{equation}
where $\psi_L^{(n)}, \, \psi_R^{(n)}$ satisfy 4D Dirac equation with mass $\mu_n$.

Multiplying the 5D Dirac equation by the conjugate of the differential operator $[ -i \Gamma^\mu \del_\mu - i \Gamma^5 \del_5 + \Phi(y)]$ and using the 4D Dirac equation for $\psi_L^{(n)}$, $\psi_R^{(n)}$, we obtain
\begin{eqnarray}
&&\left[-i\Gamma^5 \del_5 + \Phi(y) \right] \left[i\gamma^5 \del_5 + \Phi(y)\right] f_{L,R}^{(n)}\nonumber \\
&& = \left[-\del_5^2 + \Phi(y)^2 \mp \Phi'(y)\right] f_{L,R}^{(n)} = \mu_n^2 f_{L,R}^{(n)}.
\end{eqnarray} 
We can define ``creation'' and ``annihilation'' operators,
\begin{eqnarray}
a &=& \del_5 + \Phi(y), \nonumber \\
a^\dagger &=& -\del_5 +\Phi(y),
\end{eqnarray}
then we have
\begin{eqnarray}
a^\dagger a \left| f_L^{(n)} \right\rangle &=& \left[ -\del_5^2 + \Phi^2 - \Phi' \right] f_L^{(n)} = \mu_n^2 \left| f_L^{(n)} \right\rangle , \nonumber \\
a a^\dagger  \left| f_R^{(n)} \right\rangle &=& \left[ -\del_5^2 + \Phi^2 + \Phi' \right] f_R^{(n)} = \mu_n^2 \left| f_R^{(n)} \right\rangle .
\end{eqnarray}
$f_L^{(n)}, \, f_R^{(n)}$ each forms an orthonormal set. For $\mu_n^2\neq 0$, $| f_R^{(n)} \rangle = \frac{1}{\mu_n} a | f_L^{(n)} \rangle$, so $\psi_L^{(n)}, \, \psi_R^{(n)}$ are paired for nonzero modes.

Let us consider $\Phi(y) = 2\mu^2 y \propto y$, then the problem reduces to the simple harmonic oscillator. Up to a factor of $\sqrt{2} \mu$, $a,\, a^\dagger$ are just the annihilation and creation operators of a simple harmonic oscillator and the number operator $ N \propto a^\dagger a$. The states $| f_L^{(n)} \rangle, | f_R^{(n)} \rangle$ are simply
\begin{equation}
| f_L^{(n)} \rangle = | n \rangle, \quad | f_R^{(n)} \rangle = | n-1\rangle.
\end{equation}
For the zero mode,
\begin{eqnarray}
a | f_L^{(0)} \rangle =0 \quad &\Rightarrow &\quad [\del_5 + \Phi(y)] f_L^{(0)} =0,  \nonumber, \\ 
a^\dagger | f_R^{(0)} \rangle =0 \quad &\Rightarrow &\quad  [-\del_5 + \Phi(y)] f_R^{(0)} =0 .
\end{eqnarray}
The solutions are 
\begin{equation}
f_L^{(0)} (y) \propto e^{-\int_0^y \Phi(y') dy'}, \quad f_R^{(0)}(y) \propto e^{\int_0^y \Phi(y') dy'}.
\end{equation}
Only one of them is normalizable  for a domain wall: $f_L^{(0)}$ is normalizable if $\Phi(-\infty) < 0$ and $\Phi(+\infty)>0$, and $f_R^{(0)}$ is normalizable if $\Phi(-\infty) > 0$ and $\Phi(+\infty)<0$. For $\Phi(y) = 2\mu^2 y$,
\begin{equation}
f_L^{(0)} (y) =  \frac{ \mu^{1/2}}{(\pi/2)^{1/4}} e^{-\mu^2 y^2}.
\end{equation}
In this way we obtain a chiral fermion localized at $y=0$.

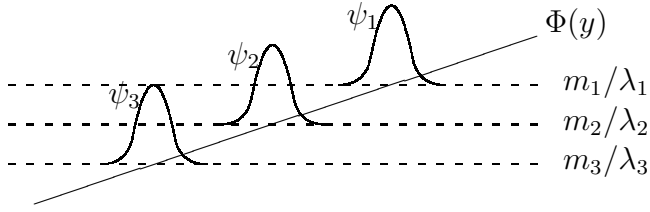
\begin{figure}
\begin{center}
\begin{picture}(250,100)(-120,-40)
\put(-90,-30){\line(3,1){190}}
\qbezier(-20,0)(-10,0)(-8,10)
\qbezier(20,0)(10,0)(8,10)
\qbezier(-8,10)(0,50)(8,10)
\qbezier(25,15)(35,15)(37,25)
\qbezier(65,15)(55,15)(53,25)
\qbezier(37,25)(45,65)(53,25)
\qbezier(-65,-15)(-55,-15)(-53,-5)
\qbezier(-25,-15)(-35,-15)(-37,-5)
\qbezier(-53,-5)(-45,35)(-37,-5)
\dashline{3}(-100,0)(100,0)
\dashline{3}(-100,15)(100,15)
\dashline{3}(-100,-15)(100,-15)
\put(110,-17){$m_3/\lambda_3$}
\put(110,-2){$m_2/\lambda_2$}
\put(110,13){$m_1/\lambda_1$}
\put(103,35){$\Phi(y)$}
\put(28,38){$\psi_1$}
\put(-17,23){$\psi_2$}
\put(-62,8){$\psi_3$}
\end{picture}
\end{center}
\caption{Chiral fermions localized at difference places in extra dimensions by a domain wall.}
\label{fig:many_fermions}
\end{figure}

One can generalize this setup to many fermion fields. Consider the action
\begin{equation}
S= \int d^4 x \,dy \sum_i \ov{\Psi} [ i\Gamma^M \del_M + \lambda_i \Phi(y) - m_i] \Psi_i .
\end{equation}
Each 5D fermion field $\Psi_i$ gives rise to a 4D chiral fermion. These chiral fermions are localized around the zeros of $\lambda_i \Phi -m_i$ (Fig.~\ref{fig:many_fermions}). 
They can be used to explain the small couplings from the small overlapping of wave functions in extra dimensions. We will consider a couple examples below.
\begin{itemize}
\item Yukawa couplings: Assuming that gauge and Higgs fields have flat wave functions in the extra dimension, we consider the following action for the lepton fields,
\begin{eqnarray}
S &=& \int d^5 x \ov{L} [ i \not \!\! \del_5  + \Phi(y) ] L + \ov{E}^c [ i \not \!\! \del_ 5 + \Phi(y) -m] E^c  \nonumber \\
&&  \qquad \qquad +(\kappa H L^T C_5 E^c + h.c.) 
\end{eqnarray} 
The zero mode of $L$ field, $l^{(0)}$ is localized at $y=0$ and the zero mode of $E^c$ field $e^{c(0)}$ is localized at $y= \frac{m}{2\mu^2} \equiv r$. The Yukawa coupling of the zero modes is then given by
\begin{eqnarray}
S_{\rm Yukawa} = \int d^4 x \,\kappa \,h(x) l(x) e^c(x) \int dy f_l^{(0)} (y) f_{e^c}^{(0)} (y),
\end{eqnarray}
where
\begin{eqnarray}
\int dy f_l^{(0)} (y) f_{e^c}^{(0)} (y) = \frac{\sqrt{2} \mu }{\sqrt{\pi}} \int dy\, e^{-\mu^2 y^2} e^{-\mu^2 (y- \frac{m}{2\mu^2})^2} = e^{-\frac{\mu^2 r^2}{2}}.
\end{eqnarray}
The Yukawa coupling is exponentially suppressed if the distance between the zero modes $r$ is somewhat bigger than $\mu^{-1}$. (See Fig.~\ref{fig:yukawa})
\begin{figure}
\begin{center}
\epsfig{file=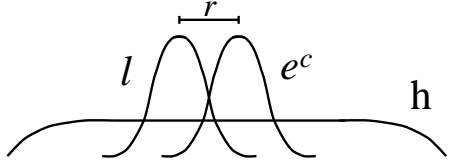,width=3.0in}
\end{center}
\caption{Suppression of lepton Yukawa coupling by split fermions.}
\label{fig:yukawa}
\end{figure}

\item Proton lifetime: The proton decay operator in 5D is 
\begin{eqnarray}
S &\sim & \int d^5 x \frac{(Q^T C_5 L)^\dagger ({U^c}^T C_5 D^c)}{M_*^3 } \nonumber \\
&\sim& \int d^4 x \frac{(q l)^\dagger (u^c d^c)}{M_*^2} \time \delta , 
\end{eqnarray}
where the suppression factor $\delta$ due to the separation of quark and lepton zero modes (Fig.~\ref{fig:proton_decay}), $r$, is
\begin{equation}
\delta \sim \int dy \left( e^{-\mu^2 y^2} \right)^3 e^{-\mu^2 (y-r)^2} \sim e^{-\frac{3}{4} \mu^2 r^2}.
\end{equation}
For $\mu r =10$, $\delta \sim 10^{-33}$ provides sufficient suppression for the proton decay rate.
\begin{figure}
\begin{center}
\epsfig{file=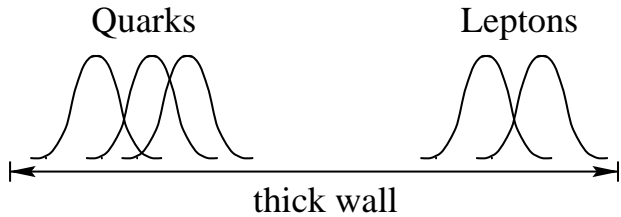,width=3.5in}
\end{center}
\caption{Suppression of proton decay by separation of quarks and leptons.}
\label{fig:proton_decay}
\end{figure}

\end{itemize}

\subsection{Boundary conditions and orbifolds}
\label{sec:orbifold}

In Sec.~\ref{sec:kk} we discussed Kaluza-Klein theory for an extra dimensin compactified on a circle (or a torus for more dimensions). To obtain realistic models we often need to compactify extra dimensions on a line segment or an orbifold in order to have a chiral theory. In this section we discuss such compactifications.

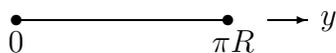
\begin{figure}
\begin{center}
\begin{picture}(150,40)(0,0)
\put(20,20){\line(1,0){80}}
\put(20,20){\circle*{4}}
\put(100,20){\circle*{4}}
\put(17,8){$0$}
\put(94,8){$\pi R$}
\put(115,20){\vector(1,0){15}}
\put(135,18){$y$}
\end{picture}
\end{center}
\caption{An extra dimension on a line segment.}
\label{fig:line}
\end{figure}

Consider a scalar field $\Phi$ living in an extra dimension which is a line segment $y \in [0, \pi R]$ (Fig.~\ref{fig:line}).
The current in $y$ directin is $J_5 = i \Phi^\dagger \del_5 \Phi$. Without addition sources at the boundaries we need $J_5$ to vanish at the boundaries to preserve unitarity. Therefore, we should have the following boundary conditions,
\begin{eqnarray}
\del_5 \Phi |_{\rm boundary} &=& 0: \qquad \mbox{Neumann boundary condition, or} \nonumber \\
\Phi |_{\rm boundary} &=& 0 : \qquad \mbox{Dirichlet boundary condition.}
\end{eqnarray}
If we impose Neumann boundary conditions at both ends, $ \del_5 \Phi (y=0) = \del_5 \Phi (y= \pi R) =0 $, The KK decomposition of $\Phi$ becomes
\begin{equation}
\Phi_+(x,y) = \frac{1}{\sqrt{\pi R}} \phi_+^{(0)} (x) + \sqrt{\frac{2}{\pi R}} \sum_{n=1}^{\infty} \phi_+^{(n)} (x) \cos \frac{ny}{R}, 
\end{equation}
with the mass $m_n=n/R$ for the $n$th mode. On the other hand , if we impose Dirichlet boundary conditions at both ends, $\Phi(y=0) = \Phi(y=\pi R)=0$, the KK decomposition of $\Phi$ is
\begin{equation}
\Phi_- (x, y) = \sqrt{\frac{2}{\pi R}} \sum_{n=1}^\infty \phi_-^{(n)} (x) \sin \frac{ny}{R}.
\end{equation}
Note that there is no zero mode for Dirichlet boundary conditions.

\begin{figure}
\begin{center}
\begin{picture}(250,60)(0,0)
\put(70,30){\circle{40}}
\put(20,30){\line(1,0){100}}
\put(50,30){\circle*{3}}
\put(90,30){\circle*{3}}
\put(70,13){\vector(0,1){34}}
\put(70,47){\vector(0,-1){34}}
\put(60,17){\vector(0,1){26}}
\put(60,43){\vector(0,-1){26}}
\put(80,17){\vector(0,1){26}}
\put(80,43){\vector(0,-1){26}}
\put(130,28){$\Rightarrow$}
\put(150,30){\line(1,0){55}}
\put(150,30){\circle*{3}}
\put(205,30){\circle*{3}}
\end{picture}
\end{center}
\caption{The $S^1/Z_2$ orbifold.}
\label{fig:orbifold}
\end{figure}
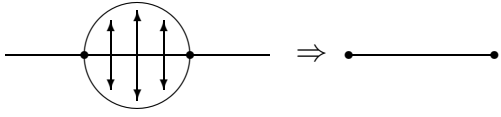

The line segment can also be described as an orbifold $S^1/Z_2$. If a theory has an exact global symmetry, we can mod out (gauge) a subgroup of the global symmetry and still obtain a consistent theory. For example, the torus compactification ($S^1$) which identifies $\Phi(y+2\pi n)$ and $\Phi(y)$ is equivalent to gauging a discrete subgroup $y\to y+2\pi n$ of translation in the $y$ direction. An orbifold is a space obtained by modding out a symmetry transformation of another space which leaves some points fixed. The simplest example is $S^1/Z_2$. Starting with a circle $S^1$, we can gauge the $Z_2$ symmetry $y \to -y$ ({\it e.g.,} identifying points $y$ and $-y$), then we obtain a line segment with fixed points $y=0$ and $y=\pi R$ (Fig.~\ref{fig:orbifold}). 

Orbifold projection requires $y \to -y$ to be a good symmetry in the original theory, so fields defined on $[-\pi R, \pi R]$ can be categorized as even or odd under $y \to -y$. They are equivalent to imposing Neumann or Dirichlet bondary conditions at the fixed points For a scalar field,
\begin{eqnarray}
\mbox{even } \Phi_+ &:& \; \Phi_+(-y) = \Phi_+ (y) \nonumber \\
&&\Rightarrow  \del_5 \Phi_+(y=0) = \del_5 \Phi_+(y=\pi R) =0, \mbox{ ``Neumann BCs,''} \nonumber \\
\mbox{odd } \Phi_- &:& \;  \Phi_-(-y)  = -\Phi_-(y) \nonumber \\
&&\Rightarrow  \Phi_-(y=0) =\Phi_-(y=\pi R)= 0, \mbox{ ``Dirichlet BCs.''}
\end{eqnarray}
We next consider a gauge field $A_M(x,y)$. Under the $Z_2$, $x_\mu \to x_\mu, \, y \to -y$, so $\del_\mu$ is even and $\del_5$ is odd. The field strength tensor $F_{\mu 5} = \del_\mu A_5 -\del_5 A_\mu$ needs to have a definite parity which implies that $A_\mu$ and $A_5$ must have opposite parities. If $A_\mu$ is even, then $A_5$ is odd and has no zero mode. Therefore in this orbifold compactification, there is no extra light scalar field from the $A_5$ component of the gauge field. Finally for the fermion field $\Psi = (\chi, \ov{\psi})^T$,
\begin{eqnarray}
S&=& \int d^5 x \; \ov{\Psi} i\Gamma^M \del_M \Psi \nonumber \\
&=& \int d^4 x\, dy\, ( i \ov{\chi} \bar{\sigma}^\mu \del_\mu \chi + i \psi \sigma^\mu \del_\mu \ov{\psi} + \psi \del_5 \chi - \ov{\chi} \del_5 \ov{\psi} ).
\end{eqnarray}
Because $\del_5$ is odd, $\psi$ and $\chi$ must have opposite parities. Only one of them has a zero mode, so the zero mode is chiral. This is also consistent with the gauge symmetry,
\begin{equation}
\ov{\Psi} \Gamma^M A_M \Psi \supset \ov{\chi} \bar{\sigma}^\mu A_\mu \chi + \psi \sigma^\mu A_\mu \ov{\psi} + \psi A_5 \chi - \ov{\chi} A_5 \ov{\psi}.
\end{equation}
On can ask whether a mass term is allowed for the fermion if $\chi$ and $\psi$ have opposite parities, $m\ov{\Psi}\Psi = m(\psi \chi + \ov{\chi}\ov{\psi})$. In the orbifold language $m$ has to be odd. However, since the theory is really just defined on one line segment, it is fine to include such a mass term. The mass term will affect the localization of the fermion zero mode as it behaves like a domain wall.

\subsection{Universal Extra Dimensions}

Standard Model fermions are chiral. We see from the previous subsection how chiral fermions arise in an orbifold compactification or from boundary conditions of extra dimensions. In such a compactification, all SM fields can propagate in the same extra dimensions. This is denoted as Universal Extra Dimensions (UEDs)~\cite{Appelquist:2000nn}.

Because all fields propagate in extra dimensions, the momentum in extra dimensions is conserved except at the fixed points (boundaries). This translates to approximate KK number conservation if the Lagrangian terms localized at the fixed points (boundaries) are ignored. As a result, tree-level contributions to the electroweak observables from KK states are suppressed. They can contribute at one-loop level. The bound on the size of UEDs is only $1/R \gtrsim 300-600$ GeV~\cite{Appelquist:2002wb,Gogoladze:2006br}.

However, boundary terms will be present as they are not forbidden by symmetries and they are induced by bulk loop corrections~\cite{Georgi:2000ks,Cheng:2002iz}. The boundary terms modify the KK spectrum from $m_{(n)} = n/R$ and lift the degeneracies of KK excitations of different SM species. Some discrete subgroup of KK number can still be preserved even including radiative corrections. For example in $S^1/Z_2$ compactification, a $Z_2$ reflection about $y= \pi R/2$ is a good symmetry if the boundary terms at $y=0$ and $y=\pi R$ are equal. (which is the case from bulk loops). As a result, the odd KK levels are odd and even KK levels are even under an exact KK-parity.  It implies that the first (odd) KK-level states have to be pair produced and the lightest first KK excitation of all SM fields is stable. It can be a good dark matter candidate if it is neutral~\cite{Servant:2002aq,Cheng:2002ej}. For collider phenomenology, the first KK states are pair-produced, then each goes through cascade decays which ends up with the stable lightest KK state. The signatures are jets/leptons plus missing energy, similar to those of SUSY with conserved $R$-parity~\cite{Cheng:2002ab}. More detailed measurements are required to distinguish these theories. 

\subsection{Symmetry breaking by orbifolds (boundary conditions)}

When there is a symmetry of the Lagrangian, the fields at the identified points in a compactification do not have to be identical, but merely equal up to a symmetry transformation. In the orbifold language, instead of modding out a discrete subgroup of the space-time symmetry, we can mod out a diagonal combination of the space-time symmetry and the internal symmetry. In this way, we can achieve symmetry breaking by orbifold compactifications. 

As an example, let us consider the breaking of the grand unified gauge group $SU(5)$ to the SM gauge group $SU(3)\times SU(2) \times U(1)$ on $S^1/Z_2$ orbifold. The embedding of $SU(3)\times SU(2) \times U(1)$ in $SU(5)$ is
\begin{equation}
SU(5) = \left( \begin{array}{ccc|cc}  &&&& \\ & SU(3) & && X,Y \\ &&&& \\ \hline &&&&\\ & X,Y&   && SU(2) \end{array} \right), \quad
U(1) \sim \begin{pmatrix} 2 &&&& \\ & 2 &&& \\ && 2 && \\ &&& -3 & \\ &&&& -3 \end{pmatrix} .
\end{equation}
Under the $Z_2$ subgroup of $SU(5)$, $\exp[ i \pi \cdot \,{\rm diag}(2,2,2,-3,-3)]$, the fundamental representation ${\bf 5}$ transforms as
\begin{equation}
{\bf 5}:  \begin{pmatrix} + \\ + \\ +\\-\\- \end{pmatrix},
\end{equation}
and the adjoint representation ${\bf 24}$ transforms as
\begin{equation}
{\bf 24} : \left( \begin{array}{ccc|cc} +&+&+&- &-\\ +& + &+&- &- \\ +&+&+&-&- \\ \hline -&-&-&+&+\\ -& - & -& +& + \end{array} \right).
\end{equation} 
Choosing the $Z_2$ of $S^1/Z_2$ orbifold to be the diagonal combination of the above $Z_2$ subgroup of $SU(5)$ and $y \to -y$, then the transformation properties of various gauge field components are
\begin{center}
\begin{tabular}{c|cc}
& $SU(3)\times SU(2) \times U(1) $ & $X,Y$ \\ \hline
$A_\mu$ & $+$ & $-$ \\
$A_5$ & $-$ & $+$\\
\end{tabular}
\end{center}
We see that only 4D gauge fields corresponding to $SU(3)\times SU(2)\times U(1)$ have zero modes, so $SU(5)$ is broken down to  $SU(3)\times SU(2)\times U(1)$~\cite{Kawamura:2000ev,Altarelli:2001qj,Hall:2001pg,Hebecker:2001jb}.

This can be described equivalently by the following boundary conditions,
\begin{center}
\begin{tabular}{c|cc}
& $SU(3)\times SU(2) \times U(1) $ & $X,Y$ \\ \hline
$A_\mu$ & (N, N) & (D, D) \\
$A_5$ & (D, D) & (N, N) \\
\end{tabular}
\end{center}
where N and D represent Neumann and Dirichelet boundary conditions respectively with first entry referring to $y=0$ and second entry referring to $y=\pi R$. In other words, while the bulk has the $SU(5)$ gauge symmetry, the boundaries only preserve $SU(3)\times SU(2)\times U(1)$.

For these boundary conditions, the $A_5$ components of the off-diagonal gauge bosons $X,Y$ have zero modes. They appear as light charged scalars in this theory. Their masses are generated from finite one-loop corrections and are of order $m \sim \frac{g}{4\pi}\frac{1}{R}$. These light scalars can be removed by choosing a different set of boundary conditions,
\begin{center}
\begin{tabular}{c|cc}
& $SU(3)\times SU(2) \times U(1) $ & $X,Y$ \\ \hline
$A_\mu$ & (N, N) & (N, D) \\
$A_5$ & (D, D) & (D, N) \\
\end{tabular}
\end{center}
In this setup, the bulk and one of the boundaries ($y=0$) preserve the full $SU(5)$ gauge symmetry, the other boundary ($y=\pi R$) only preserves $SU(3)\times SU(2)\times U(1)$. In the orbifold language, this corresponds to an $S^1/( Z_2 \times Z'_2)$ orbifold.


We have seen that an orbifold compactification can be equivalently described by boundary conditions. In fact, we can also consider more general boundary conditions~\cite{Hebecker:2001jb,Csaki:2003dt}. Let us start with a bulk action for the scalar field
\begin{equation}
S_{bulk}=\int d^4x \, dy \int_0^{\pi R} \left( \frac{1}{2}\, \partial^M
\phi \,
\partial_M \phi -V(\phi )\right),
\end{equation}
Consider first for simplicity that there is no boundary term. Applying the variation principle we obtain
\begin{eqnarray}
\delta S &=& \int d^4x \, dy \int_0^{\pi R} \left(
\partial^M\phi \partial_M \delta \phi -
\frac{\partial V}{\partial \phi} \delta \phi \right) \nonumber \\
&=& \int d^4x \int_0^{\pi R} dy \left[ -\partial_\mu
\partial^\mu \phi \delta \phi -\frac{\partial V}{\partial \phi}
\delta \phi -\partial_y \phi
\partial_y \delta \phi \right] \nonumber \\
&=&\int d^4x \int_0^{\pi R} dy \left[ -\partial_M \partial^M \phi
 -\frac{\partial V}{\partial \phi} \right] \delta \phi
-\left[ \int d^4 x \partial_y \phi \delta \phi \right]_0^{\pi R}.
\end{eqnarray}
For $\delta S=0$ we require that the field satisfies the bulk equation of motion (EOM) 
\begin{equation}
\partial_M \partial^M \phi
= -\frac{\partial V}{\partial \phi}
\end{equation} 
and also the boundary variation needs to vanish,
\begin{equation}
\partial_y \phi \delta \phi\vert_{\rm boundary} =0.
\label{scalarBC}
\end{equation}
This can be satisfied by the Neumann boundary condition $\del_y \phi \vert_{\rm boundary}=0$ for arbitrary
$\delta \phi\vert_{\rm boundary}$. We call it natural boundary condition if $\delta \phi\vert_{\rm boundary}$ is left free to vary.
 
Other boundary conditions can be obtained by adding boundary terms. For example, we can include boundary mass terms,
\begin{equation}
S=S_{\rm bulk}-\int d^4x \frac{1}{2} M_1^2 \phi^2|_{y=0}-\int d^4x
\frac{1}{2} M_2^2 \phi^2|_{y=\pi R}.
\end{equation}
Variation of the action gives
\begin{eqnarray}
\delta S_{\rm boundary}&=& \int d^5 x \mbox{ (EOM) } \delta \phi \nonumber \\
&& -\int \delta \phi (\partial_y \phi +M_2^2 \phi )|_{y=\pi R}
+\int d^4 x \delta \phi (\partial_y \phi -M_1^2 \phi)|_{y=0}.
\end{eqnarray}
The natural BC's will be given by
\begin{eqnarray}
&& \partial_y \phi +M_2^2 \phi =0 \ \ {\rm at} \ \ y=\pi R, \nonumber \\
&& \partial_y \phi -M_1^2 \phi =0 \ \ {\rm at} \ \ y=0.
\end{eqnarray}
If we take $M_1,\, M_2  \to \infty$, then $\phi =0$ at $y=0, \, \pi R$ and $\del_y \phi$ will be arbitrary at $y=\epsilon, \, \pi R -\epsilon$, infinitesimally away from the boundaries. This is equivalent to the Dirichlet boundary conditions. We can always understand the Dirichlet boundary conditions
as the case with infinitely large boundary mass terms for the fields.

Similarly for a gauge field,
\begin{equation}
S=\int d^5x (-\frac{1}{4} F_{MN}^a F^{MN\,a} )= \int d^5
x(-\frac{1}{4} F_{\mu \nu}^a F^{\mu\nu\,a}-\frac{1}{2} F_{\mu 5}^a
F^{\mu 5\,a}),
\end{equation}
the natural boundary conditions in unitary gauge are $\del_y A_\mu^a =0$, $A_5^a=0$.

Now we add boundary scalar fields which have nonzero VEVs,
\begin{equation}
{\cal L}_{i}=|D_\mu \Phi_i|^2 -\lambda_i (|\Phi_i|^2-\frac{1}{2}v_i^2)^2,
\end{equation}
where $i=1$ is at $y=0$ and $i=2$ is at $y=\pi R$.
These boundary terms will induce non-vanishing VEVs and $\Phi_i$ can be parametrized as a physical Higgs and a
Goldstone boson,
\begin{equation}
\Phi_i=\frac{1}{\sqrt{2}} (v_i+h_i) e^{i \pi_i/v_i}.
\end{equation}
In this case, the natural boundary conditions become
\begin{equation}
\partial_y A_\mu \mp v_{1,2}^2 A_\mu \vert_{1,2}=0.
\end{equation}
Taking $v_i \to \infty$, we obtain $A_\mu \vert_{1,2}=0$, which corresponds to Dirichlet bondary conditions. The Higgs and the Goldstone $h_i$, $\pi_i$ decouple from the gauge field in this limit. At the same time, $A_5$ boundary conditions change from Dirichlet to Neumann boundary conditions $\del_y A_5=0$. Note that the mass of the lightest gauge boson in this limit is $\sim 1/R$, independent of $v_i$. 

Gauge symmetry breaking can be achieved by choosing appropriate boundary conditions. For $SU(2) \to U(1)$, we can set $A_\mu^{1,2}(y=0) = A_\mu^{1,2}(y=\pi R) =0$, $\del_y A_\mu^3 (y=0) = \del_y A_\mu^3 (y=\pi R) =0$. This is equivalent to orbifold breaking. However, we can also reduce the rank of the gauge group by breaking $SU(2)$ to nothing, which can not be achieved by a simple orbifold. We just need to choose $A_\mu^{1,2}(y=0) =0,\, \del_y A_\mu^3 (y=0) =0$, and $A_\mu^{2,3}(y=\pi R) =0,\, \del_y A_\mu^1(y=\pi R) =0$, then no gauge field has a zero mode and the gauge symmetry is completely broken.

The discussion of boundary conditions for fermions can be found in Ref.~\refcite{Csaki:2003sh}. One can build realistic models with electroweak symmetry broken by boundary conditions~\cite{Csaki:2003sh,Csaki:2003zu}.

\section{Dimension Deconstruction}
\label{sec:deconstruction}

Gauge theories in more than 4 dimensions are non-renormalizable. They are treated as low energy effective theories below some cutoff $\Lambda$. A na\"{i}ve momentum cutoff breaks gauge invariance. It also violates locality in extra dimensions. We need to be careful about how to implement the cutoff and asking cutoff-sensitive questions, otherwise we can easily get non-sensible answers if we do not regularize the theories in a correct way. One way to regularize the higher dimensional gauge theories while preserving gauge invariance and locality is to put extra dimensions on a lattice~\cite{ArkaniHamed:2001ca,Hill:2000mu,Cheng:2001vd}.

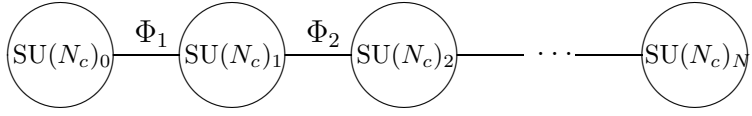
\begin{figure}
\begin{center}
\begin{picture}(300,50)(0,0)
\put(30,25){\circle{40}}
\put(50,25){\line(1,0){25}}
\put(12,22){\footnotesize SU$(N_c)_0$}
\put(58,30){$\Phi_1$}
\put(95,25){\circle{40}}
\put(77,22){\footnotesize SU$(N_c)_1$}
\put(115,25){\line(1,0){25}}
\put(123,30){$\Phi_2$}
\put(160,25){\circle{40}}
\put(142,22){\footnotesize SU$(N_c)_2$}
\put(180,25){\line(1,0){25}}
\put(210,25){$\ldots$}
\put(225,25){\line(1,0){25}}
\put(270,25){\circle{40}}
\put(251,22){\footnotesize SU$(N_c)_N$}
\end{picture}
\end{center}
\caption{A lattice of $SU(N_c)$ gauge groups.}
\label{fig:moose}
\end{figure}

Consider $N+1$ copies of $SU(N_c)$ gauge groups in 4D, with $N$ link-Higgs fields $\Phi_i$ which transform as bi-fundamentals $(N_{c,i}, \ov{N}_{c,i-1})$ under neighboring gauge groups (Fig.~\ref{fig:moose}),
\begin{equation}
{\cal{L}}_{QCD}= -\frac{1}{4}\sum_{i=0}^N F_{i\mu\nu}^a F^{i\mu\nu a} +
\sum_{i=1}^{N} D_{\mu}\Phi_i^\dagger D^{\mu}\Phi_i ,
\end{equation}
\begin{equation}
D_{\mu}= \partial_{\mu} + i
\tilde{g} \sum_{i=0}^{N} 
A_{i\mu}^{a}T_i^{a} ,
\end{equation}
where $\tilde{g}$ is the gauge coupling of $SU(N_c)_i$ group (assumed to be identical for simplicity). We can write down a potential for each link-Higgs field such that each $\Phi_i$ develops a VEV of the form
\begin{equation}
\langle \Phi_i \rangle = v \cdot {\bf I}
\end{equation}
where ${\bf I}$ is the $(N+1)\times (N+1)$ identity matrix. We assume that all left-over physical Higgs bosons are heavy (with mass $> v$).  

These VEVs break $N+1$ $SU(N_c)$ gauge groups down to the diagonal $SU(N_c)$ gauge group. The mass matrix for the $N+1$ sets of gauge fields $A_{i\mu}^a, \, (i =0, \cdots, N)$ is
\begin{equation}
\label{mat1}
M =   \frac{1}{2}{\tilde{g}^2v^2}\left(
\begin{array}{ccccc}
1&-1&0&\cdots&0 \\
-1&2&-1& \cdots&0 \\
0&-1 &2 &\cdots&0 \\
\vdots& \vdots & & \cdots & \\
0 & 0 & \cdots & 2 & -1 \\
0&0&\cdots&-1&1
\end{array} \right).
\end{equation}
The mass eigenstates $\tilde{A}_\mu^n$ can be obtained by diagonalizing the mass matrix,
\begin{equation}
A_{\mu}^j = \sum_{n=0}^{N} a_{jn} \tilde{A}_{\mu}^n. 
\label{AA}
\end{equation}
For $n \neq 0$,
\begin{equation}
a_{jn} =\sqrt{\frac{2}{N+1}} \cos{(\frac{2j+1}{2}\gamma_n})\ , \qquad
j = 0, 1,\dots, N, 
\label{anj}
\end{equation}
where $\gamma_n=\pi n/(N+1)$, and 
\begin{equation}
a_{j0} = \frac{1}{\sqrt{N+1}}\, \qquad j = 0, 1,\dots, N .
\end{equation}
The masses for $\tilde{A}_\mu^n$ are given by 
\begin{equation}
\label{Mn}
M_n = {2}\tilde{g} v \sin \left( \frac{\gamma_n}{2} \right)
= {2}\tilde{g} v \sin \left( \frac{n\pi}{2(N+1)} \right) , \qquad n=0,1,\dots, N. 
\end{equation}
For small $n$,
\begin{equation}
M_n \approx   \frac{\tilde{g}v\pi n}{(N+1)},\qquad \qquad n \ll N .
\end{equation}
They are the same as the KK tower from a compactified extra dimension
if we identify
\begin{equation}
\frac{\tilde{g}v\pi}{(N+1)} = \frac{1}{R} = \frac{\pi}{L} \quad \Rightarrow \quad L = \frac{N+1}{\tilde{g} v}.  
\end{equation}
This corresponds to compactification on a line segment (or $S^1/Z_2$ orbifold). The low energy gauge coupling of the remaining diagonal gauge group is given by
\begin{equation}
g=\tilde{g}/\sqrt{N+1}.
\end{equation}

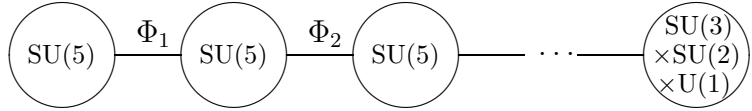
\begin{figure}
\begin{center}
\begin{picture}(300,50)(10,0)
\put(30,25){\circle{40}}
\put(50,25){\line(1,0){25}}
\put(17,22){\footnotesize SU(5)}
\put(58,30){$\Phi_1$}
\put(95,25){\circle{40}}
\put(82,22){\footnotesize SU(5)}
\put(115,25){\line(1,0){25}}
\put(123,30){$\Phi_2$}
\put(160,25){\circle{40}}
\put(147,22){\footnotesize SU(5)}
\put(180,25){\line(1,0){25}}
\put(210,25){$\ldots$}
\put(225,25){\line(1,0){25}}
\put(270,25){\circle{40}}
\put(258,33){\footnotesize SU(3)}
\put(253,22){\footnotesize $\times$SU(2)}
\put(255,11){\footnotesize $\times$U(1)}
\end{picture}
\end{center}
\caption{The deconstructed version of orbifold symmetry breaking.}
\label{fig:deconstruction}
\end{figure}

The 4D theory with $N+1$ $SU(N_c)$ gauge groups provides a renormalizable UV completion of 5D gauge theory with a cutoff 
\begin{equation}
\Lambda \sim \frac{1}{a} = \frac{1}{\tilde{g} v},
\end{equation}
where $a$ is the lattice spacing. Many results in higher-dimensional theories can be easily translated into the 4D language with this dimension-deconstruction. For example, the orbifold symmetry breaking of $SU(5) \to SU(3)\times SU(2) \times U(1)$ can be realized by considering many copies of $SU(5)$ gauge groups with one copy of $SU(3)\times SU(2)\times U(1)$ gauge group at the end in 4D, broken by the link fields to the diagonal  $SU(3)\times SU(2)\times U(1)$ gauge group (Fig.~\ref{fig:deconstruction})~\cite{Csaki:2001qm,Cheng:2001qp}.

For gauge theories, there is one-to-one correspondence between the higher-dimensional theory and a 4D theory with a ``theory space.'' However, dimension-deconstruction for gravity is more difficult, there are strong-coupling issues involving the scalar longitudinal degrees of freedom and the cutoff is much lower than the na\"ively expected value~\cite{ArkaniHamed:2002sp}. It is beyond the scope of this lecture. Interested readers are referred to the original paper of this discussion.

\section{Epilogue}

In these lectures I tried to give a brief summary of theories with flat extra dimensions developed in recent years. The goal is to give a jump start for advanced graduate students who are interested in research in this area. The topics covered here are certainly not complete, but just a sample of what people have been thinking of in these directions. Hopefully they provide enough basics so that the students are able to find and understand the other related subjects in the literature.
There are also many other useful review articles and lectures on extra dimensions with different emphases which are good resources for further explorations~\cite{Rubakov:2001kp,Gabadadze:2003ii,Csaki:2004ay,PerezLorenzana:2005iv,Sundrum:2005jf,Csaki:2005vy,Rattazzi:2003ea}.

\section*{Acknowledgments}
I would like to thank Csaba Cs\'{a}ki for organizing the TASI 2009 summer school and inviting me to give the lectures.
This work is supported in part by the Department of Energy Grant DE-FG02-91ER40674.

\end{document}